\documentclass[twoside]{ASDJ}

\usepackage{epsfig}
\usepackage[outercaption]{sidecap}
\usepackage{hyperref} 
\usepackage{tikz}
\usepackage{cite}
\usepackage{amsmath}
\hypersetup{colorlinks=true} 
\usepackage{caption}

\usepackage[T3,T1]{fontenc}
\DeclareSymbolFont{tipa}{T3}{cmr}{m}{n}
\DeclareMathAccent{\invbreve}{\mathalpha}{tipa}{16} 

\begin{document}

\ASDJvolume{7}
\ASDJnumber{1}
\ASDJyear{2019}
\ASDJpage{1}
\ASDJreceived{November 1, 2019}
\ASDJrevised{Dec 10, 2019}
\ASDJaccepted{Dec 29, 2019}
\ASDJarticle{53}

\author{\mbox{Marcello} Righi$^1$, Marco Berci$^2$}

\affiliations{$^1$ Zurich University of Applied Sciences (ZHAW) and Swiss Federal Institute of Technology Zurich (ETHZ). \\ $^2$ Pilatus Aircraft Ltd.}

\shortauthor{M. Righi, M. Berci}
\shorttitle{On Elliptical Wings in Subsonic Flow}

\title{On Elliptical Wings in Subsonic Flow: Indicial Lift Generation via CFD Simulations - with Parametric Analytical Approximations}

\maketitle

\begin{abstract}
This study deals with the process and benefits of generating aerodynamic indicial-admittance functions with the support of computational fluid dynamics based on the Euler equations. A simple parametric analytical approximation is alongside developed based on linear lifting line and piston theory, for the sake of assessing the physical consistency of numerical solutions while investigating the suitability of theoretical models in the subsonic regime. Thin elliptical wings are hence considered and the design space is described by three parameters: wing aspect ratio (6, 8, 12, 20),  free-stream Mach number (0.3, 0.5, 0.6) and flow perturbation type (angle of attack step, vertical sharp-edged gust). All numerical simulations exhibit good quality and weak sensitivity to parameters other than  Mach number and flow aspect ratio, accounting for their combined effects. The benefit of using numerical results to tune analytical approximations and improve their accuracy is finally demonstrated by matching both initial and final behaviours of the wing lift development, with an effective synthesis of the high-fidelity features in the low-fidelity model.
\end{abstract}








\section{Introduction}
\label{sec:introduction}

Multidisciplinary design and optimisation (MDO) \cite{Alexandrov} 
benefits from robust and efficient methods for aeroelastic stability investigation and dynamic loads calculation \cite{Kier2}. To this end, indicial-admittance functions \cite{Leishman1} may serve as very effective tools and reduced-order models (ROMs) for characterising unsteady airloads \cite{Raveh,Gennaretti}; more detailed literature may be found in previous works \cite{Berci-AST-2017,Righi,righi2016subsonic,DARONCH2018617}.

In order to assess the derivation process of such functions, one might limit the analysis to the subsonic regime \cite{Wright}. Over the years, a few exact and approximate indicial functions have respectively been obtained for the lift build-up of thin aerofoil \cite{Wagner,Kussner,VonKarman} and finite wing \cite{Jones1,jones1945,queijo1978approximate} in both incompressible and compressible flow \cite{Lomax1,Lomax2}. These functions include both circulatory and non-circulatory parts, the physical phenomena behind which can be approached either numerically via computational fluid dynamics (CFD) \cite{Parameswaran} or analytically via mathematical models such as potential flow \cite{Pike}. CFD-based generation of aerodynamic indicial functions for thin aerofoil and finite wing has been performed by numerous researchers in recent years \cite{Silva,Cavagna2,Romanelli,DaRonch,Ghoreyshi1,Ghoreyshi2,Jansson}. Time-accurate schemes, physically consistent grid motion and deformation algorithms \cite{Cizmas} have reached a good maturity level \cite{Farhat1,Farhat2} and are implemented in the most popular CFD solvers developed for the aeronautical community; however, for the accurate simulation of both circulatory and non-circulatory portions of the flow response there is no established best practice yet. Thus, this work extends and improves the accuracy of previous studies by the authors concerning the derivation of indicial lift functions for thin wings in subsonic flow \cite{Righi,righi2016subsonic,DARONCH2018617}.

The present study focuses on elliptical thin wings and three parameters are considered: wing aspect ratio $\eta$ (6, 8, 12, 20),  free-stream Mach number $M$ (0.3, 0.5, 0.6) and flow perturbation type (unit angle of attack step "AoA", unit vertical sharp-edged gust "SEG"); the wing section is a NACA0002 aerofoil for all cases. CFD simulations are based on Euler equations and benefit from rigorous studies on the effects of domain size, spatial resolution, time stepping and integration scheme. A convenient parametric expression is then proposed for approximating indicial lift functions of compressible subsonic flow by modifying those of incompressible flow, based on the Prandtl-Glauert scalability for the circulatory part and piston theory for the non-circulatory part \cite{Beddoes,Berci-AST-2017}. In order to mimic the CFD solutions \cite{Wieseman,Berci3}, the analytical expression then uses two values from the latter: initial rate and final value of the lift build-up, within a consistent framework.

This paper is structured as follows: Section \ref{sec:num} describes the process followed in order to generate accurate CFD simulations, Section \ref{sec:analytical} presents the analytical approximations method, the comparison of numerical and analytical results is then discussed in Section \ref{sec:discussion} and conclusions are drawn in Section \ref{sec:conclusions}. All computational grids and configuration files for the CFD solver SU2 as well as post-processing scripts for Scilab are made available on request to all interested parties.

\section{Numerical Investigations}
\label{sec:num}

Setting up and running CFD solutions has followed a conventional process; however, a few considerations are reported here. 

\subsection{CFD equations and solver}
\label{subsec:solver}

In the present study, the Euler equations are solved using the open-source Stanford University Unstructured (SU2) code \cite{economon2016su2}. The finite volume method is applied on arbitrary unstructured meshes using a standard edge-based data structure on a dual grid with control volumes constructed using a median-dual, vertex-based scheme. Regarding time integration, SU2 is capable to solve implicitly steady and unsteady problems, using a dual-time stepping strategy\cite{jameson2009}, achieving second-order accuracy in space and time. Even though the choice may appear surprising in a work aiming for accuracy, the conventional Roe's scheme (notoriously dissipative) has been used for all simulations; yet, in these simulations Roe's scheme has demonstrated to minimise numerical oscillations and no efforts have been done in order to reduce or optimise the numerical dissipation of the convective scheme.

Note that the Navier-Stokes equations could alternatively have been chosen with suitably high Reynolds number. Since fine grids have also been used for the Euler equations, viscous calculations would have requested a manageable increase in computational resources (say, an increase of 100\%) and provided a solution much less prone to numerical oscillations; however, the dependence on both Reynolds number and settings of the turbulence model would have been impossible to eliminate, while the present study focuses on aerodynamic phenomena which are essentially inviscid. Provided the boundary layer thickness remains limited, no significant difference is expected to be added by a viscous solution \cite{Berci-AST-2017}.

\subsection{Perturbations generation and flow response linearity}
\label{subsec:input}

The linearity of the solution can be appreciated in a relatively large range of angle of attack (i.e., over 4 degrees) and is shown in Section \ref{sec:discussion}. The magnitude of the flow perturbation has then been chosen to comfortably lie inside the linear region while still being much larger than numerical noise: one degree of angle of attack for all test cases except those at $M=0.6$, for which the magnitude has been halved in order to avoid the appearance of regions of supersonic flow. 

The angle of attack step is generated by adding a fictitious grid velocity field normal to the wing plane (i.e., by letting the computational domain move vertically at constant speed). This approach has worked without any particular problems; however, the same approach does not work as smoothly to simulate a sharp-edged gust. Even though this is a well-accepted technique used to introduce a gust in many CFD solvers, we noticed it prone to develop small numerical oscillations at the beginning of the unsteady simulation when the gust is sharp-edged; of course, the oscillations are much smaller when the gust shape is smooth (e.g., for a classic 1-cos profile). Such numerical oscillations have been observed with more than one solver and are hence not concerned with SU2 specifically. 

In order to provide an accurate response to a sharp-edged gust advancing across the wing, a simple yet effective alternative approach has been used: a perturbation field was added to the steady-state solution at the beginning of the unsteady run. The perturbation consists of a uniform velocity field, including a velocity component normal to the wing plane, which is added upstream of the wing. By doing so, the main flow carries the perturbation field across the wing, creating the same effect as the one due to a gust moving across. With a suitable magnitude (the same used for the angle of attack step change) and distance from the wing leading-edge, the method has proven very effective and surprisingly accurate; the most suitable distance has proven to be about $0.01$ chords, allowing the gust front to maintain its sharpness before impacting the wing.
{
Since the vertical gust is introduced as a physical perturbation and not as a perturbation grid-velocity field, this approach accounts for the influence of the wing geometry on the gust shape. The deterioration of the travelling gust shape caused by numerical diffusion and dispersion was already characterised and found acceptable for practical applications \cite{Basu,Berci3}.}

\subsection{Spatial and temporal resolutions}
\label{subsec:resolution}

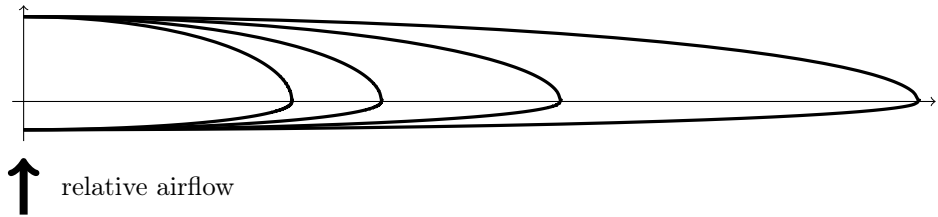
\begin{figure}

\begin{center}

\begin{tikzpicture}[scale=0.75]

 \draw [very thick,rounded corners=1 pt,color=black] (0.000000e+00, -1.000000e+00) -- 
(4.605521e-01, -9.976064e-01) -- 
(9.128034e-01, -9.905301e-01) -- 
(1.349041e+00, -9.790737e-01) -- 
(1.762641e+00, -9.637055e-01) -- 
(2.148419e+00, -9.450133e-01) -- 
(2.502798e+00, -9.236513e-01) -- 
(2.823801e+00, -9.002887e-01) -- 
(3.110898e+00, -8.755655e-01) -- 
(3.364766e+00, -8.500599e-01) -- 
(3.587001e+00, -8.242672e-01) -- 
(3.779839e+00, -7.985904e-01) -- 
(3.945890e+00, -7.733382e-01) -- 
(4.087937e+00, -7.487297e-01) -- 
(4.208765e+00, -7.249016e-01) -- 
(4.311051e+00, -7.019156e-01) -- 
(4.397290e+00, -6.797645e-01) -- 
(4.469749e+00, -6.583733e-01) -- 
(4.530455e+00, -6.375904e-01) -- 
(4.581192e+00, -6.171606e-01) -- 
(4.623512e+00, -5.966501e-01) -- 
(4.658751e+00, -5.752251e-01) -- 
(4.688052e+00, -5.507496e-01) -- 
(4.712389e+00, -5.000000e-01) -- 
(4.712389e+00, -5.000000e-01) -- 
(4.688052e+00, -3.477511e-01) -- 
(4.658751e+00, -2.743246e-01) -- 
(4.623512e+00, -2.100498e-01) -- 
(4.581192e+00, -1.485181e-01) -- 
(4.530455e+00, -8.722871e-02) -- 
(4.469749e+00, -2.488007e-02) -- 
(4.397290e+00, 3.929365e-02) --  
(4.311051e+00, 1.057467e-01) -- 
(4.208765e+00, 1.747047e-01) -- 
(4.087937e+00, 2.461892e-01) -- 
(3.945890e+00, 3.200146e-01) -- 
(3.779839e+00, 3.957711e-01) -- 
(3.587001e+00, 4.728016e-01) -- 
(3.364766e+00, 5.501796e-01) -- 
(3.110898e+00, 6.266966e-01) -- 
(2.823801e+00, 7.008662e-01) -- 
(2.502798e+00, 7.709540e-01) -- 
(2.148419e+00, 8.350399e-01) -- 
(1.762641e+00, 8.911165e-01) -- 
(1.349041e+00, 9.372211e-01) -- 
(9.128034e-01, 9.715904e-01) -- 
(4.605521e-01, 9.928191e-01) -- 
(0.000000e+00, 1.000000e+00);

   \draw [very thick,rounded corners=1 pt,color=black] (0.000000e+00, -1.000000e+00) -- 
(6.140695e-01, -9.976064e-01) -- 
(1.217071e+00, -9.905301e-01) -- 
(1.798721e+00, -9.790737e-01) -- 
(2.350188e+00, -9.637055e-01) -- 
(2.864558e+00, -9.450133e-01) -- 
(3.337065e+00, -9.236513e-01) -- 
(3.765068e+00, -9.002887e-01) -- 
(4.147864e+00, -8.755655e-01) -- 
(4.486354e+00, -8.500599e-01) -- 
(4.782669e+00, -8.242672e-01) -- 
(5.039785e+00, -7.985904e-01) -- 
(5.261187e+00, -7.733382e-01) -- 
(5.450583e+00, -7.487297e-01) -- 
(5.611687e+00, -7.249016e-01) -- 
(5.748069e+00, -7.019156e-01) -- 
(5.863053e+00, -6.797645e-01) -- 
(5.959665e+00, -6.583733e-01) -- 
(6.040607e+00, -6.375904e-01) -- 
(6.108257e+00, -6.171606e-01) -- 
(6.164682e+00, -5.966501e-01) -- 
(6.211667e+00, -5.752251e-01) -- 
(6.250737e+00, -5.507496e-01) -- 
(6.283185e+00, -5.000000e-01) -- 
(6.283185e+00, -5.000000e-01) -- 
(6.250737e+00, -3.477511e-01) -- 
(6.211667e+00, -2.743246e-01) -- 
(6.164682e+00, -2.100498e-01) -- 
(6.108257e+00, -1.485181e-01) -- 
(6.040607e+00, -8.722871e-02) -- 
(5.959665e+00, -2.488007e-02) -- 
(5.863053e+00, 3.929365e-02) -- 
(5.748069e+00, 1.057467e-01) -- 
(5.611687e+00, 1.747047e-01) -- 
(5.450583e+00, 2.461892e-01) -- 
(5.261187e+00, 3.200146e-01) -- 
(5.039785e+00, 3.957711e-01) -- 
(4.782669e+00, 4.728016e-01) -- 
(4.486354e+00, 5.501796e-01) -- 
(4.147864e+00, 6.266966e-01) -- 
(3.765068e+00, 7.008662e-01) -- 
(3.337065e+00, 7.709540e-01) -- 
(2.864558e+00, 8.350399e-01) -- 
(2.350188e+00, 8.911165e-01) -- 
(1.798721e+00, 9.372211e-01) -- 
(1.217071e+00, 9.715904e-01) -- 
(6.140695e-01, 9.928191e-01) -- 
(0.000000e+00, 1.000000e+00);

   \draw [very thick,rounded corners=1 pt,color=black] (0.000000e+00, -1.000000e+00) -- 
(9.211043e-01, -9.976064e-01) -- 
(1.825607e+00, -9.905301e-01) -- 
(2.698082e+00, -9.790737e-01) -- 
(3.525281e+00, -9.637055e-01) -- 
(4.296838e+00, -9.450133e-01) -- 
(5.005597e+00, -9.236513e-01) -- 
(5.647603e+00, -9.002887e-01) -- 
(6.221796e+00, -8.755655e-01) -- 
(6.729531e+00, -8.500599e-01) -- 
(7.174003e+00, -8.242672e-01) -- 
(7.559677e+00, -7.985904e-01) -- 
(7.891781e+00, -7.733382e-01) -- 
(8.175874e+00, -7.487297e-01) -- 
(8.417530e+00, -7.249016e-01) -- 
(8.622103e+00, -7.019156e-01) -- 
(8.794580e+00, -6.797645e-01) -- 
(8.939498e+00, -6.583733e-01) -- 
(9.060911e+00, -6.375904e-01) -- 
(9.162385e+00, -6.171606e-01) -- 
(9.247024e+00, -5.966501e-01) -- 
(9.317501e+00, -5.752251e-01) -- 
(9.376105e+00, -5.507496e-01) -- 
(9.424778e+00, -5.000000e-01) -- 
(9.424778e+00, -5.000000e-01) -- 
(9.376105e+00, -3.477511e-01) -- 
(9.317501e+00, -2.743246e-01) -- 
(9.247024e+00, -2.100498e-01) -- 
(9.162385e+00, -1.485181e-01) -- 
(9.060911e+00, -8.722871e-02) -- 
(8.939498e+00, -2.488007e-02) -- 
(8.794580e+00, 3.929365e-02) -- 
(8.622103e+00, 1.057467e-01) -- 
(8.417530e+00, 1.747047e-01) -- 
(8.175874e+00, 2.461892e-01) -- 
(7.891781e+00, 3.200146e-01) -- 
(7.559677e+00, 3.957711e-01) -- 
(7.174003e+00, 4.728016e-01) -- 
(6.729531e+00, 5.501796e-01) -- 
(6.221796e+00, 6.266966e-01) -- 
(5.647603e+00, 7.008662e-01) -- 
(5.005597e+00, 7.709540e-01) -- 
(4.296838e+00, 8.350399e-01) -- 
(3.525281e+00, 8.911165e-01) -- 
(2.698082e+00, 9.372211e-01) -- 
(1.825607e+00, 9.715904e-01) -- 
(9.211043e-01, 9.928191e-01) -- 
(0.000000e+00, 1.000000e+00);

   \draw [very thick,rounded corners=1 pt,color=black] (0.000000e+00, -1.000000e+00) -- 
(1.535174e+00, -9.976064e-01) -- 
(3.042678e+00, -9.905301e-01) -- 
(4.496803e+00, -9.790737e-01) -- 
(5.875469e+00, -9.637055e-01) -- 
(7.161396e+00, -9.450133e-01) -- 
(8.342661e+00, -9.236513e-01) -- 
(9.412671e+00, -9.002887e-01) -- 
(1.036966e+01, -8.755655e-01) -- 
(1.121589e+01, -8.500599e-01) -- 
(1.195667e+01, -8.242672e-01) -- 
(1.259946e+01, -7.985904e-01) -- 
(1.315297e+01, -7.733382e-01) -- 
(1.362646e+01, -7.487297e-01) -- 
(1.402922e+01, -7.249016e-01) -- 
(1.437017e+01, -7.019156e-01) -- 
(1.465763e+01, -6.797645e-01) -- 
(1.489916e+01, -6.583733e-01) -- 
(1.510152e+01, -6.375904e-01) -- 
(1.527064e+01, -6.171606e-01) -- 
(1.541171e+01, -5.966501e-01) -- 
(1.552917e+01, -5.752251e-01) -- 
(1.562684e+01, -5.507496e-01) -- 
(1.570796e+01, -5.000000e-01) -- 
(1.570796e+01, -5.000000e-01) -- 
(1.562684e+01, -3.477511e-01) -- 
(1.552917e+01, -2.743246e-01) -- 
(1.541171e+01, -2.100498e-01) -- 
(1.527064e+01, -1.485181e-01) -- 
(1.510152e+01, -8.722871e-02) -- 
(1.489916e+01, -2.488007e-02) -- 
(1.465763e+01, 3.929365e-02) -- 
(1.437017e+01, 1.057467e-01) -- 
(1.402922e+01, 1.747047e-01) -- 
(1.362646e+01, 2.461892e-01) -- 
(1.315297e+01, 3.200146e-01) -- 
(1.259946e+01, 3.957711e-01) -- 
(1.195667e+01, 4.728016e-01) -- 
(1.121589e+01, 5.501796e-01) -- 
(1.036966e+01, 6.266966e-01) -- 
(9.412671e+00, 7.008662e-01) -- 
(8.342661e+00, 7.709540e-01) -- 
(7.161396e+00, 8.350399e-01) -- 
(5.875469e+00, 8.911165e-01) -- 
(4.496803e+00, 9.372211e-01) -- 
(3.042678e+00, 9.715904e-01) -- 
(1.535174e+00, 9.928191e-01) -- 
(0.000000e+00, 1.000000e+00);

\draw [<-,color=black] (0,1.2) -- (0,-1.2); 
\draw [->,color=black] (-0.2,-0.5) -- (16,-0.5); 

\draw [->,color=black,line width=3pt] (0,-2.5) -- (0,-1.5); 
\draw (0.5,-2) node[anchor=west] {relative airflow};
\end{tikzpicture}

\end{center}

\caption{Half-wing geometry for all test cases.}
\label{fig:planform}
\end{figure}

\begin{figure}
\begin{center}
\includegraphics[width=0.45\textwidth]{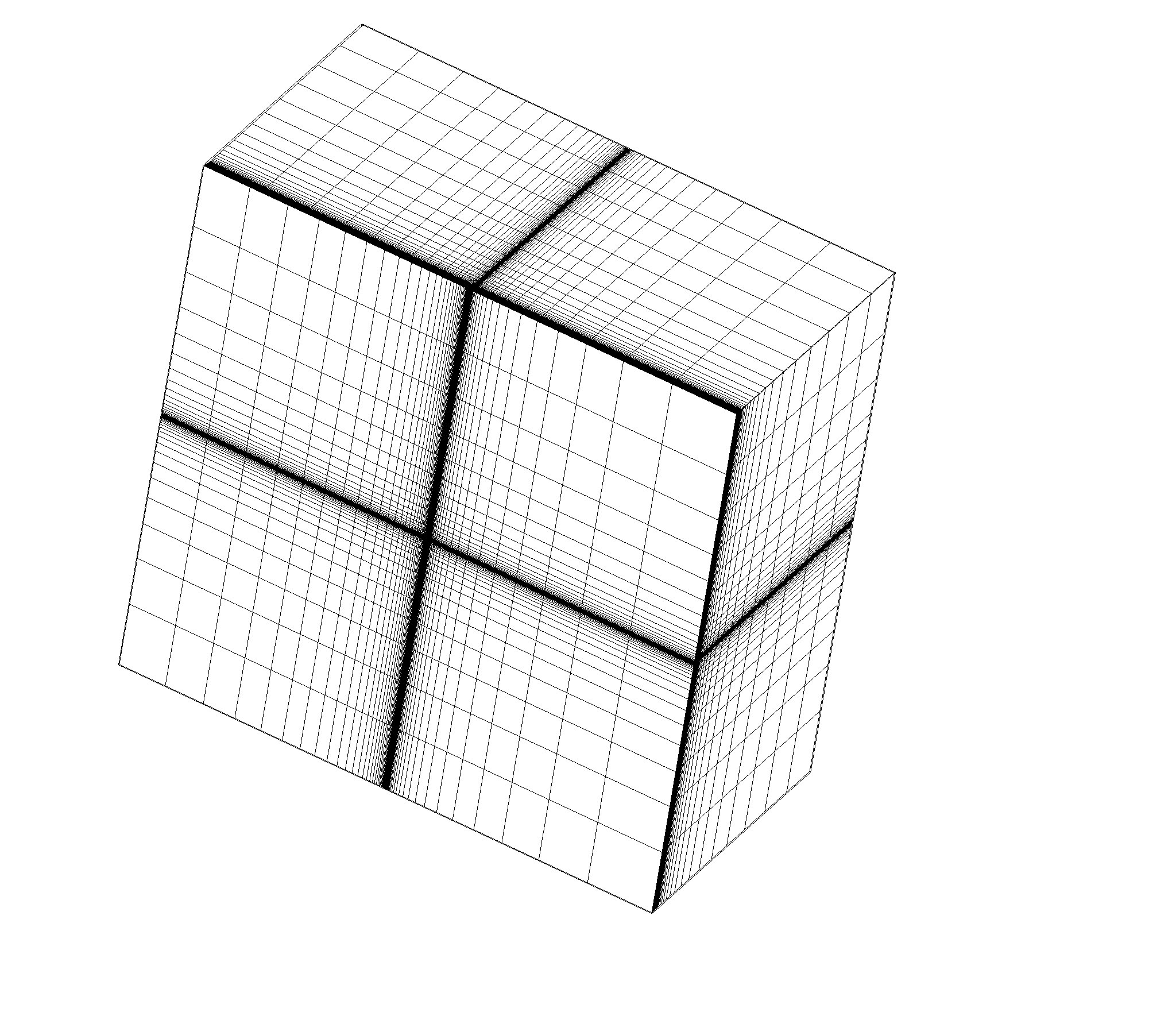}
\includegraphics[width=0.45\textwidth]{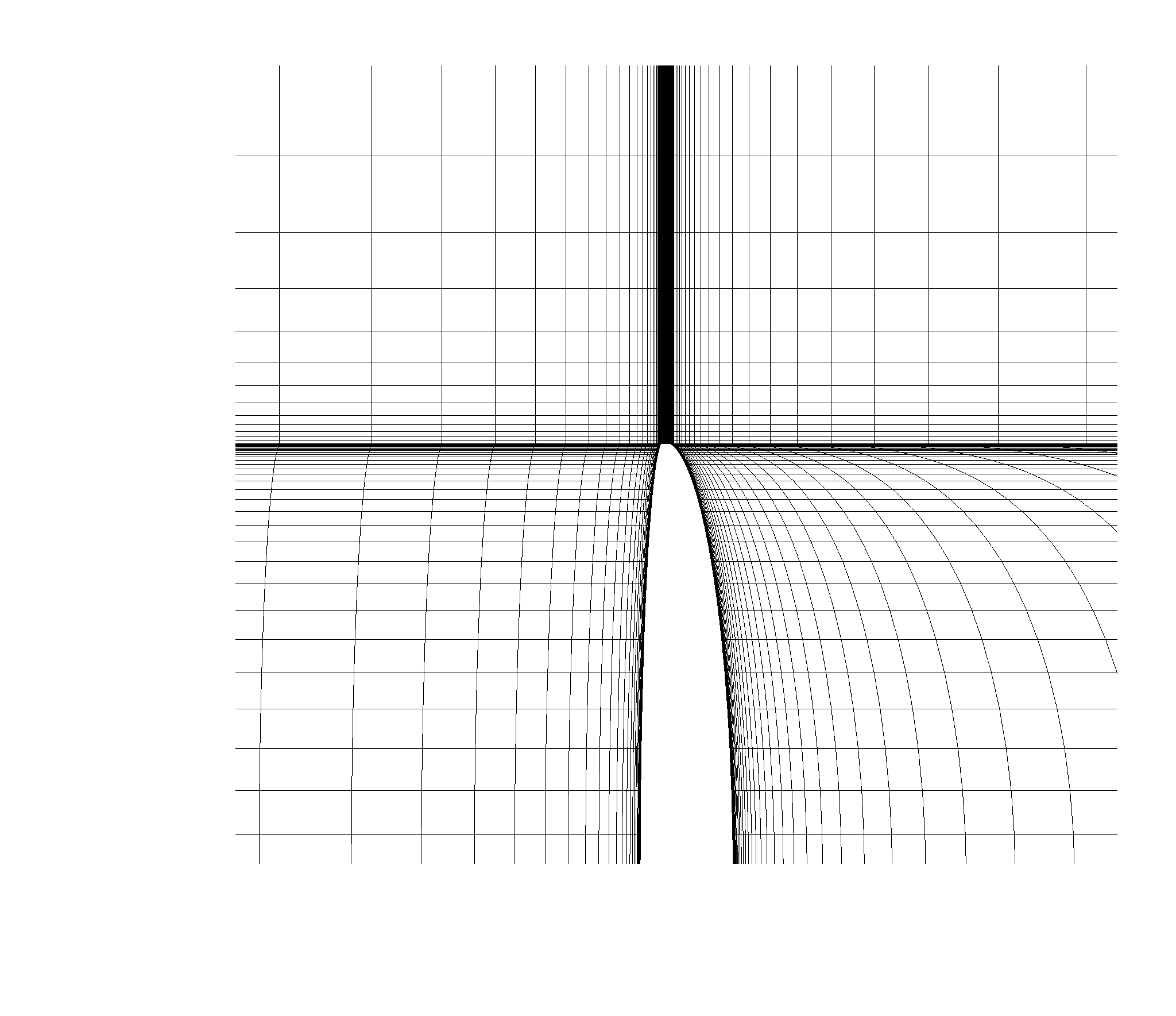}
\end{center}
\caption{Three-dimensional view of the computational domain and section for $z=0$ and $\eta=12$.}
\label{fig:domainandgrid}
\end{figure}

The wing geometry is shown in Fig. \ref{fig:planform}. The computational domains are rectangular boxes and contain the left half-wing only (see Figure \ref{fig:domainandgrid}), taking of advantage of the problem symmetry. The size of the domains, $400\times 400\times 200$ chords in $x$, $z$ and $y$, has been fixed in order to prevent pressure waves generated by the interaction between flow perturbation and boundary conditions (i.e., "slip" on wet surfaces and "far-field" on free surfaces) from affecting the flow around the wing; additional details may be found in previous publications \cite{Righi,Berci-AST-2017,DARONCH2018617,righi2016subsonic}.

Particular care has been taken in the generation of the computational grids, which are entirely structured and composed of multiple blocks; mesh files in SU2 format have been generated with a relatively simple script written in Scilab (\url{https://wiki.scilab.org/}). All grids are structured (H-H) and built with great attention to quality. Despite the use of Euler equations, the grid cells are stretched near the wing surface almost like a grid for the Navier-Stokes equations, with a cell size the order of $10^{-5}$ chords; this is done to provide a higher accuracy in the reconstruction of the pressure gradients due to the wing motion as well as to minimise the numerical oscillations (a similar study based on conventional unstructured computational grids and conventional spacing did show substantial numerical oscillations \cite{Righi}). 

Grids size and spacing have been fixed on the basis of several convergence studies carried out for various configurations and Mach numbers; in the end, 192 cells have been placed on either side of each wing section, 54 cells upstream and 54 cells downstream. The sizes of the computational cells on the leading and trailing edge are of the order of $10^{-4}$ chords.
{
The cells spacing normal to the wing surface has been stretched more than in a conventional Euler grid, with cell height in the order of $10^{-5}$ chords. In the course of previous studies \cite{Righi,DARONCH2018617}, the authors have assessed the sensitivity of the results to the spatial resolution normal to the wing surface; a positive correlation between quality of CFD results (or absence of spurious numerical oscillations) and the stretching ratio normal to the surface has been  consistently noticed. We believe that a finer grid in the vicinity of the wall may resolve in  a more robust way the strong pressure gradients appearing in the impulsive part of the response, even if no boundary layer must be resolved.  


}

\begin{figure}
\begin{center}
\includegraphics[width=0.30\textwidth]{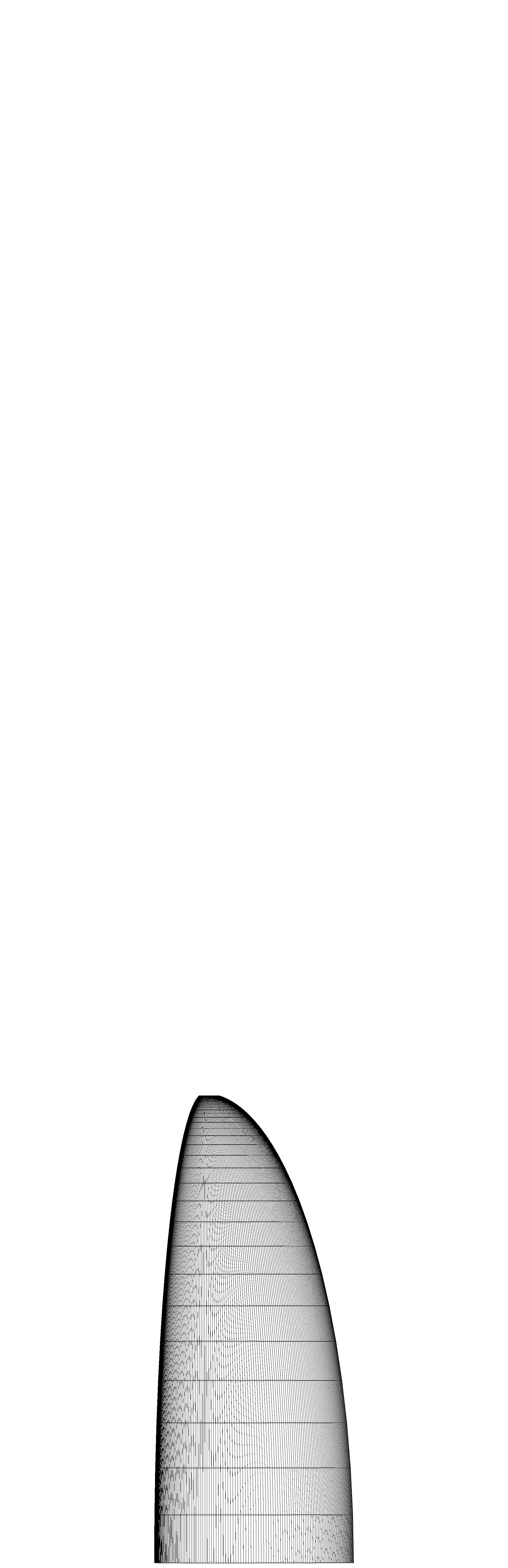} \hspace{-12mm}
\includegraphics[width=0.30\textwidth]{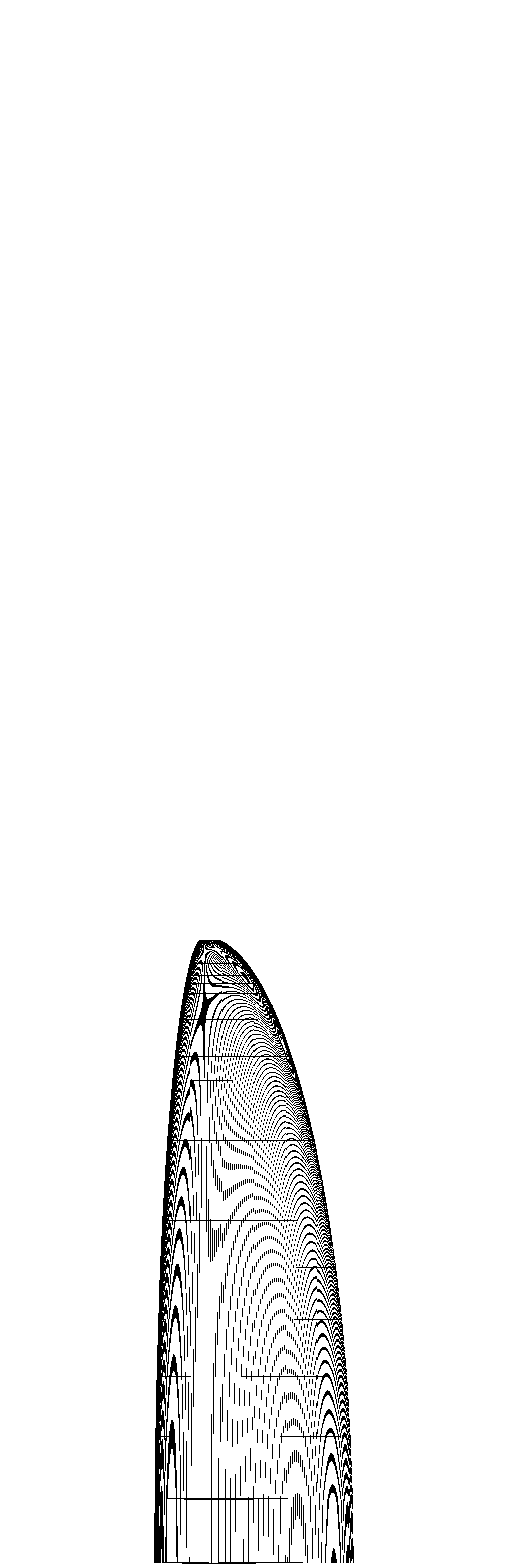} \hspace{-12mm}
\includegraphics[width=0.30\textwidth]{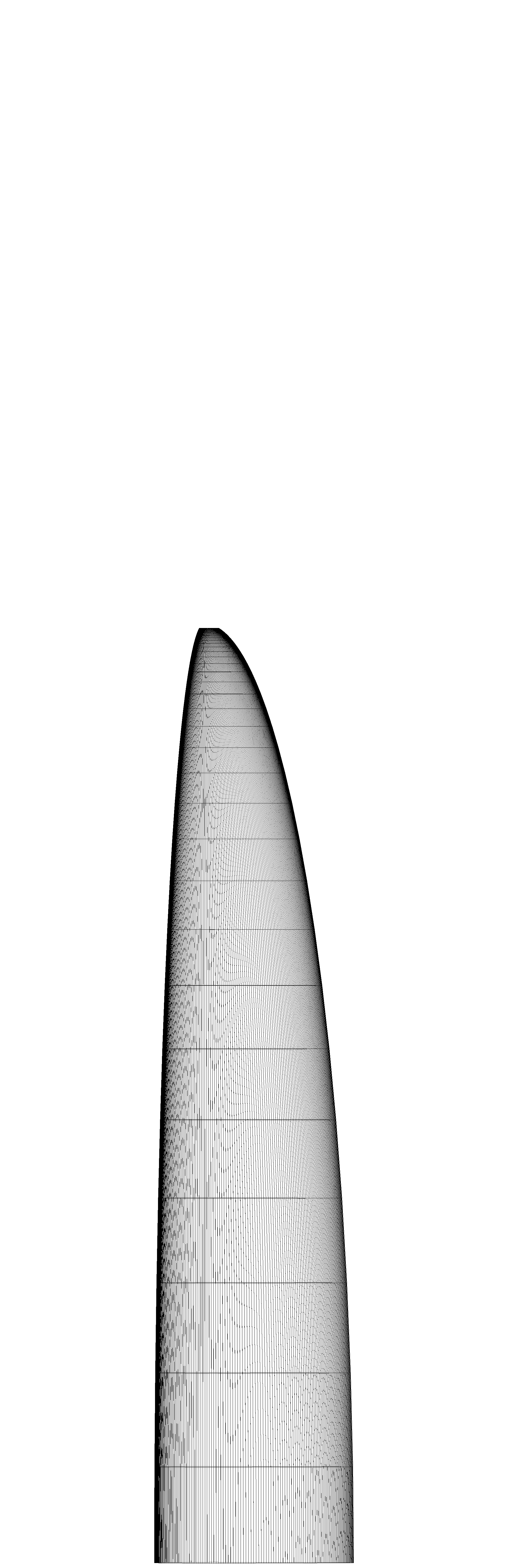} \hspace{-12mm}
\includegraphics[width=0.30\textwidth]{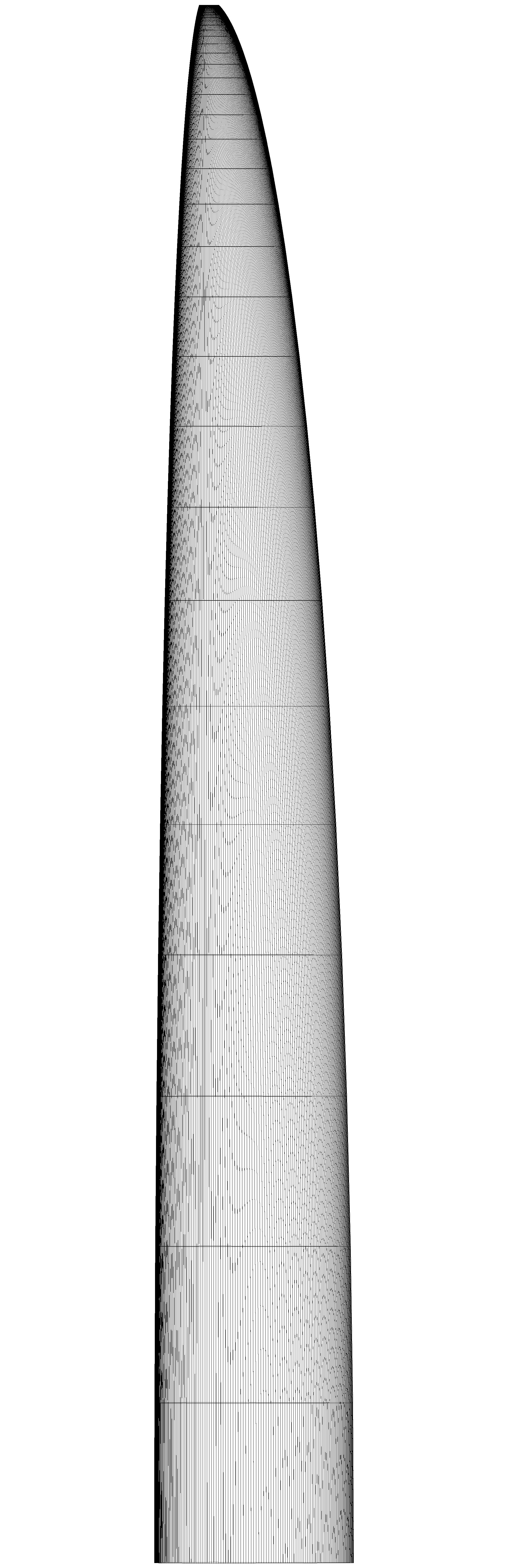}
\end{center}
\caption{Surface mesh for all four Aspect Ratio values. }
\label{fig:ar12grid}
\end{figure}

As for the $y$ dimension, 36 cells have been allocated between the symmetry plane and the wing tip plane, whereas additional 28 cells have been used between the tip planes and the outer boundary; several tests have been carried out with a higher resolution, without substantial improvements. The surface grids used are shown in Figure \ref{fig:ar12grid}.

The flow response has been obtained by integrating the Euler equations in time. As in the case of two-dimensional analyses \cite{Berci-AST-2017}, the temporal resolution has been subject of different studies which have provided the value $\Delta\tau=0.02$ for $M=0.3$ as the maximum reduced-time step size for the impulsive part of the response. In order to save on computational time, the time step is subsequently increased by 10 times after 400 iterations (i.e. at a reduced time $\tau=8$ for $M=0.3$).

\section{Analytical Approximations}
\label{sec:analytical}

The present approximation of the indicial functions for compressible subsonic flow was proposed in previous works \cite{Righi} and is obtained in the reduced time $\tau$ from the corresponding incompressible one \cite{Jones1,Berci4}, by means of the Prandtl-Glauert transformation for the circulatory part and piston theory for the non-circulatory part \cite{Leishman1}, which are then linearly superposed \cite{Berci-AST-2017}. With respect to the lift development for a unit step in the angle-of-attack and a unit sharp-edged gust, analytical expressions for the circulatory contribution are available for finite wings in both incompressible \cite{jones1945,queijo1978approximate} and compressible flow \cite{Jones3,DARONCH2018617}. As far as the the non-circulatory contribution is concerned, analytical solutions are available for thin aerofoils only \cite{Lomax1,Lomax2}, whereas numerical solutions shall be obtained for finite wings \cite{Miranda}. The proposed analytical model is then tuned to match the limit behaviour of the CFD simulations \cite{Wieseman,Palacios} in both circulatory and non-circulatory parts, with $\breve{k}(\eta,M)$ and $\invbreve{k}(\eta,M)$ tuning parameters for the final value and initial rate of change of the wing lift, which are respectively taken from the (steady) asymptotic and (unsteady) impulsive-like numerical results. The initial value of circulatory and non-circulatory contributions coincide in both tuned and untuned cases \cite{Pike}; then, the unsteady lift develops with a different rate, with two-dimensional theory applicable for infinitely slender wing \cite{Hernandes}.


With $l$ the semi-span and $\bar{c}$ the root chord, elliptical wings own surface $S=\frac{\pi}{2} l \bar{c}$ and aspect ratio $\eta=\frac{8 l}{\pi \bar{c}}$. Prandtl's lifting line theory \cite{Prandtl} is a powerful method for calculating the steady lift distribution of slender straight wings in potential subsonic flow \cite{Berci-Aerospace}; however, it is known slightly conservative by neglecting second-order terms in the governing equation for the wing circulation \cite{Diederich,Bisplinghoff} and the model accuracy increases with increasing the wing's aspect ratio \cite{Jones2}. 

Assuming the same aerofoil and reference angle-of-attack $\alpha$ for all sections $0 \leq y \leq l$, elliptical wings own geometric chord $c(y)$, circulation $\bar{\Gamma}(y)$ and lift coefficient $\bar{C}_L$ generally given by:
\begin{equation}
c = \bar{c} \sqrt{1-\left(\frac{y}{l}\right)^2},   \qquad \qquad
\bar{\Gamma} = \frac{U}{2} c \bar{C}_L,   \qquad \qquad
\bar{C}_L = \frac{\pi \eta \breve{k} \bar{C}_l}{\pi \eta + \bar{C}_{l/\alpha}},
\end{equation}
respectively, where the lift coefficient $\bar{C}_l$ and its derivative $\bar{C}_{l/\alpha}$ for the isolated wing section include most of the thickness and compressibility effects already; in particular, $\bar{C}_l=\bar{C}_{l/\alpha}\alpha$ and $\bar{C}_{l/\alpha}=\frac{2\pi}{\beta}$ for a flat airfoil, with $\beta=\sqrt{1-M^2}$ the compressibility factor \cite{Glauert2,Gothert}. Note that the downwash angle $\bar{\alpha}_i=\frac{\bar{C}_L}{\pi \eta}$ is constant along the span \cite{Katz} and all wing sections experience the same effective angle-of-attack $\alpha-\bar{\alpha}_i$ as well as pressure distribution along the chord \cite{Karamcheti}.


Aiming at a simple parametric approximation of the unsteady lift coefficient $C_L(\tau)$, two exponential terms reproduce the circulatory $\breve{C}_L(\tau)$ and non-circulatory $\invbreve{C}_L(\tau)$ contributions, namely \cite{Leishman1,Righi}:
\begin{equation}
C_L = \bar{C}_L \left( 1 - \breve{A} e^{-\breve{B} \beta^2 \tau} \right) + \invbreve{A} e^{-\invbreve{B} \beta^2 \tau},  \qquad \qquad
\tau=\left(\frac{2U}{\bar{c}}\right)t,
\end{equation}
where the coefficients $\breve{A}$, $\invbreve{A}$ and $\breve{B}$, $\invbreve{B}$ are found by imposing the initial and asymptotic behaviours of the numerical simulations \cite{Leishman4}, which involve the relative tuning parameters $\breve{k}$ and $\invbreve{k}$ directly (see the Appendix). No attempt has intentionally been made to replicate the lift behaviour in the transitional region from non-circulatory to circulatory flow response by curve-fitting the CFD results, within a conservative approach for practical applications \cite{Berci-AST-2017,Beddoes} (where the circulatory low-frequency part is typically more relevant than the non-circulatory high-frequency part \cite{Venkatesan,Bellinger}).

Enforcing the limit behaviours for a unit step in the angle of attack gives \cite{Righi}:
\begin{align}
\breve{A}^{\perp} &= 1-\cfrac{\pi}{E \bar{C}_L},   \\
\breve{B}^{\perp} &= \cfrac{1}{4E} \left[\cfrac{2+\eta}{\left(2E-1\right)\eta-2}\right],  \\
\invbreve{A}^{\perp} &= \cfrac{4}{M}-\cfrac{\pi}{E}, \\
\invbreve{B}^{\perp} &= \cfrac{E M}{4E-\pi M} \left\{ \left(\cfrac{\bar{C}_L}{4E}-\cfrac{\pi}{4E^2}\right) \left[ \cfrac{2+\eta}{\left(2E-1\right)\eta-2} \right]
                     + 2\invbreve{k}^{\perp} \left(\cfrac{1-M}{M^2 \beta^2}\right) \right\},
\end{align}
whereas for a unit vertical sharp-edged gust it is \cite{Righi}:
\begin{align}
\breve{A}^{\dashv} &= \left(1-\frac{\pi}{E \bar{C}_L}\right)\sigma,   \\
\breve{B}^{\dashv} &= \cfrac{1}{4E} \left[\cfrac{2+\eta}{\left(2E-1\right)\eta-2}\right],   \\  
\invbreve{A}^{\dashv} &= \left(\bar{C}_L-\frac{\pi}{E}\right)\sigma - \bar{C}_L, \\
\invbreve{B}^{\dashv} &= \cfrac{E}{\left(E\bar{C}_L-\pi\right)\sigma-E\bar{C}_L} \left\{ \left(\cfrac{\bar{C}_L}{4E}-\cfrac{\pi}{4E^2}\right) 
                        \left[ \cfrac{\left(2+\eta\right)\sigma}{\left(2E-1\right)\eta-2} \right]-\cfrac{2\invbreve{k}^{\parallel}}{\sqrt{M} \beta^2} \right\},
\end{align}
where the parameter $\sigma$ is defined from reciprocal relations \cite{Garrick} as:
\begin{equation}
\sigma = e^{\cfrac{9\beta^2}{32E} \left[\cfrac{2+\eta}{\left(2E-1\right)\eta-2}\right]},
\end{equation}
and introduces the net penetration effect on the circulatory lift build-up \cite{Mazelsky1}; approximations for two-dimensional potential flow are then retrieved for infinitely slender wings. For the sake of comparison, Table \ref{tab:gainspoles} shows the circulatory coefficients for a unit step in the angle of attack of flat elliptical plates in incompressible potential flow as numerically obtained by previous authors \cite{Jones1}.

\begin{table}
\begin{center}
\begin{tabular}{l|ccc}
$\eta$ & 3 & 6 & $\infty$ \\ \hline 
$E$ & 1.165 & 1.055 & 1.000 \\ \hline
$\bar{C}_L$ & 3.770 & 4.712 & 6.283 \\
$\breve{A}^{\perp}$ & 0.285 & 0.368 & 0.500 \\
$\breve{B}^{\perp}$ & 0.539 & 0.406 & 0.250 \\ \hline
$\bar{C}_L$ \cite{Jones1} & 3.770 & 4.712 & 6.283 \\ 
$\breve{A}^{\perp}$ \cite{Jones1} & 0.283 & 0.361 & 0.500 \\ 
$\breve{B}^{\perp}$ \cite{Jones1} & 0.540 & 0.381 & 0.215 \\ 
\end{tabular}
\end{center}
\caption{Circulatory coefficients for a unit step in the angle of attack of a flat elliptical plate in incompressible potential flow}
\label{tab:gainspoles}
\end{table}

\section{Results Discussion and Cross-Validation}
\label{sec:discussion}

All analysed test cases are listed in Table \ref{tab:tc} and include several combinations of different parameters: four wing aspect ratios, three free-stream Mach numbers, two flow perturbation types; a total of 24 test cases have then been computed.

\begin{table}
\begin{center}
\begin{tabular}{l|c|c}
Aspect ratio & Mach number & Flow perturbation \\ \hline 
6, 8, 12, 20 & 0.3, 0.5, 0.6 & AoA, SEG  \\
\end{tabular}
\end{center}
\caption{Test cases analysed.}
\label{tab:tc}
\end{table}

\subsection{CFD quality check}
\label{subsec:quality}

\begin{figure}
    \centering
    \includegraphics[width=0.49\textwidth]{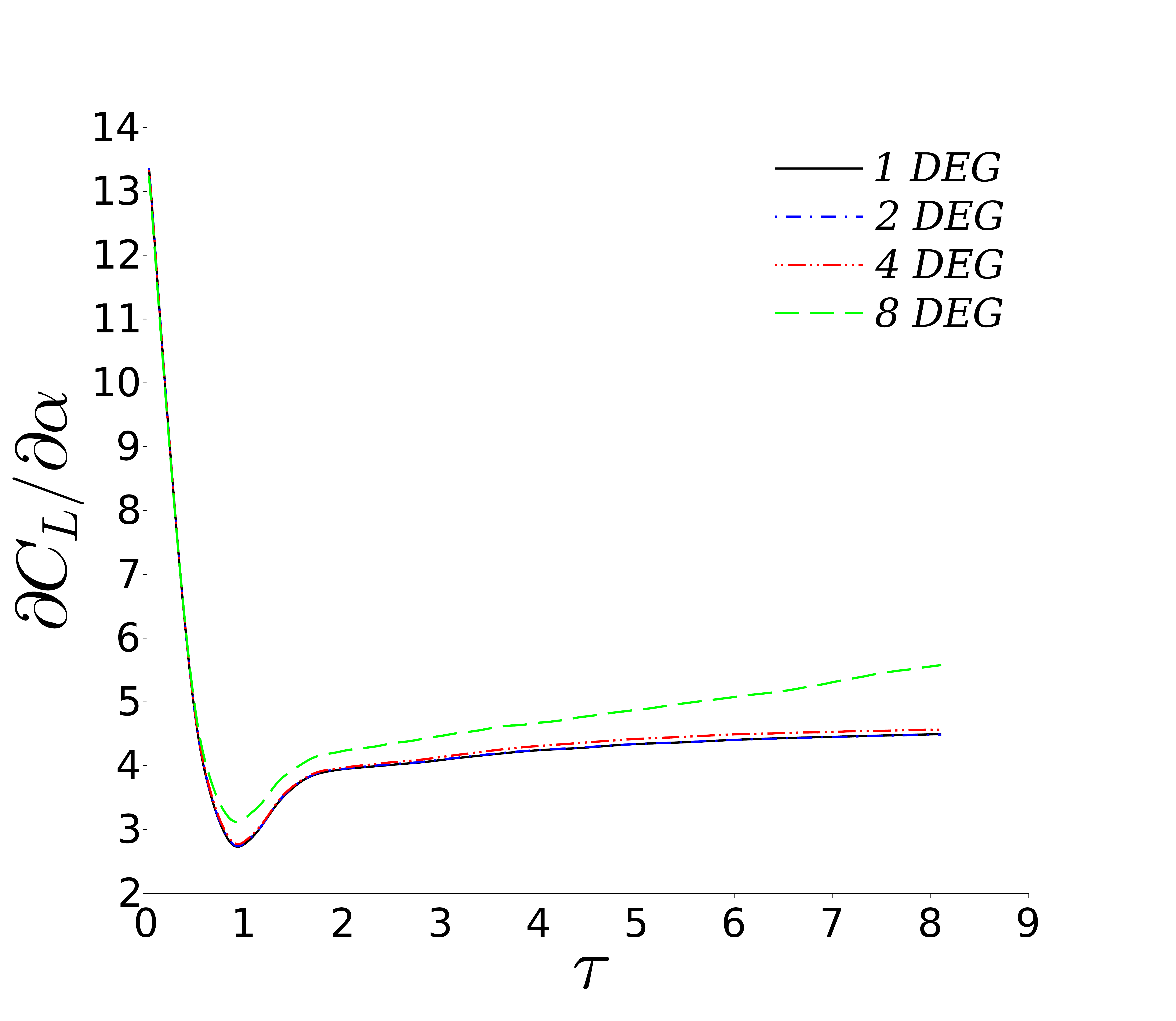}
    \includegraphics[width=0.49\textwidth]{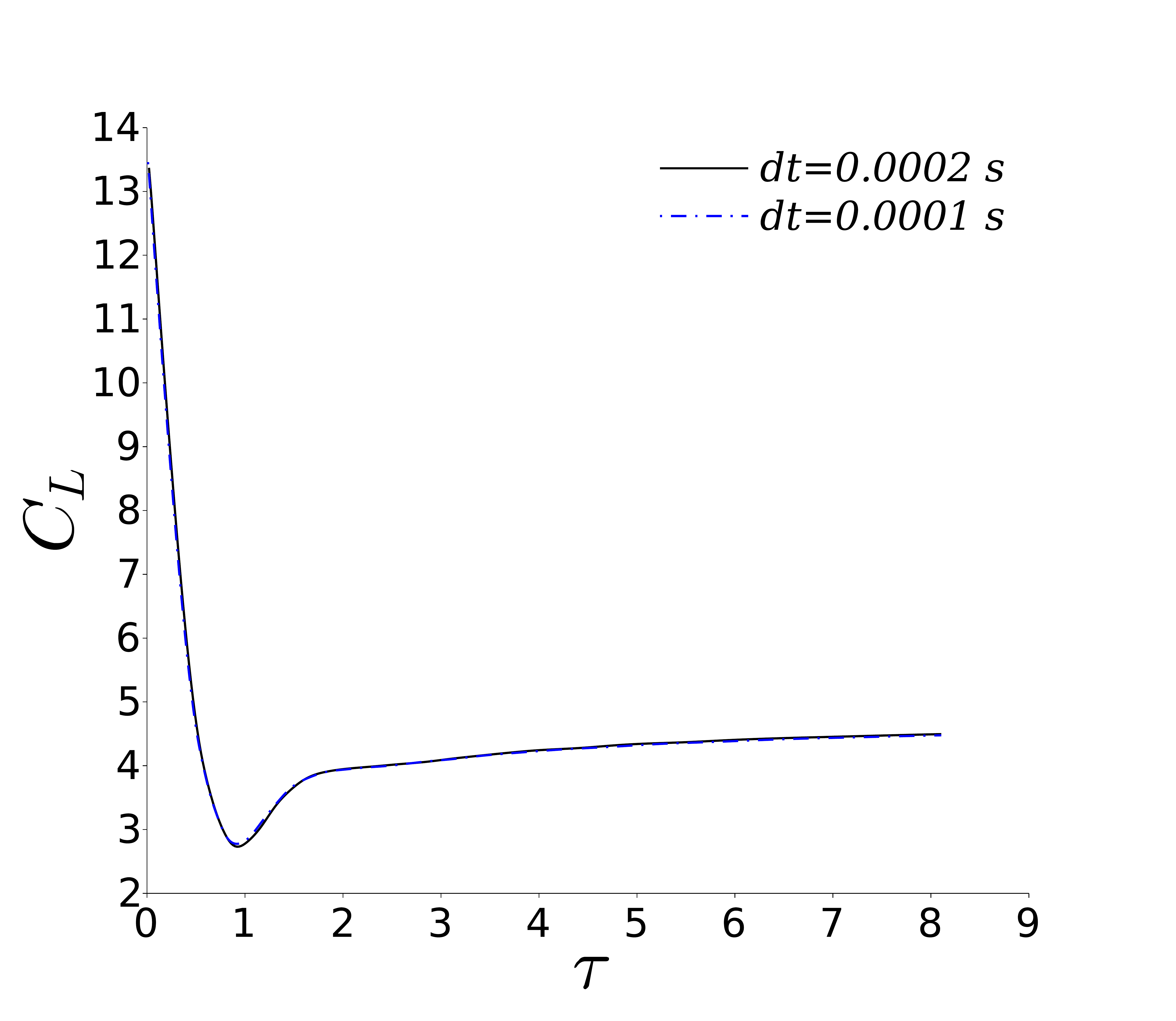}
    \caption{Indicial response to AoA step with different flow perturbation magnitude (left) and time-step length (right), for $M=0.3$ and $\eta=6$.}
    \label{fig:convergence}
\end{figure}

Adequacy of grid spacing has been thoroughly checked in the previous studies \cite{Berci-Aerospace}. Adequacy of the magnitude of the flow perturbation is shown in Figure \ref{fig:convergence}, where the flow response is reasonably linear with a reference angle of attack up to 4 degrees; thus, all subsequent simulations used either 1 or 0.5 degrees. Adequacy of time step length is also shown in Figure \ref{fig:convergence}; note that halving the time step causes a maximum error of 1\% only in the transitional region of the flow response.

\subsection{CFD results}
\label{subsec:results}

\begin{figure}
    \centering
    \includegraphics[width=0.49\textwidth]{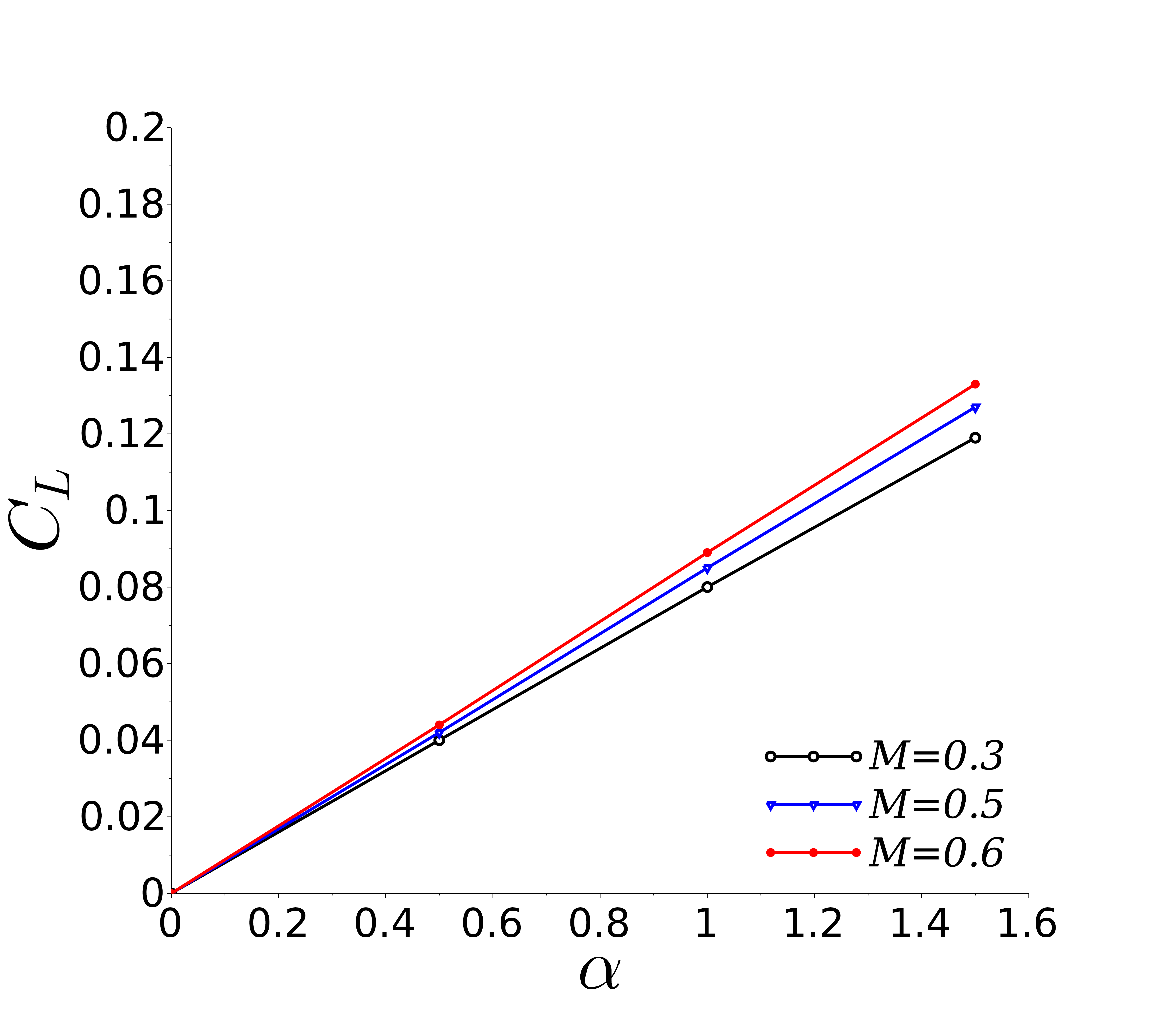}
    \includegraphics[width=0.49\textwidth]{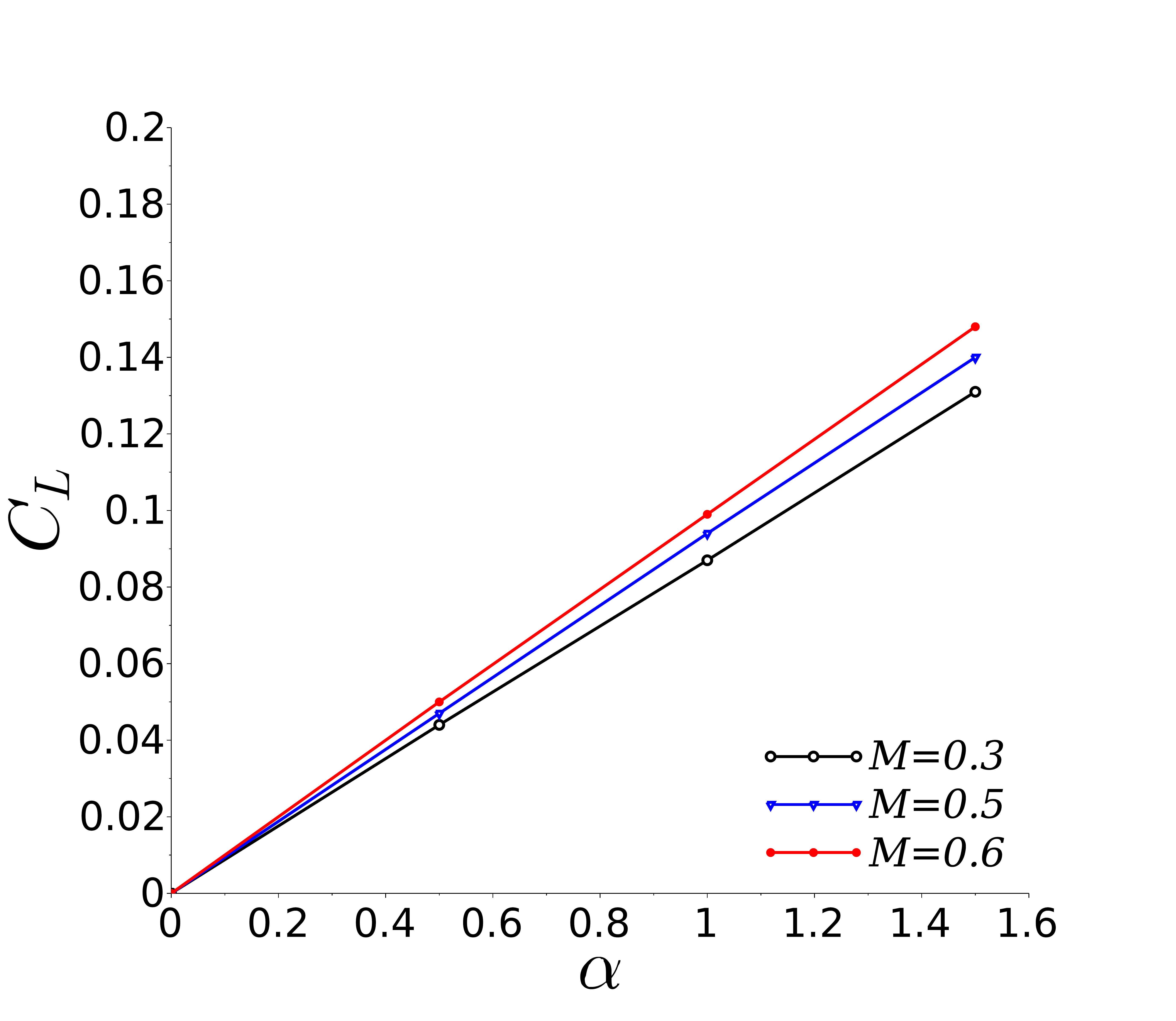}
    \includegraphics[width=0.49\textwidth]{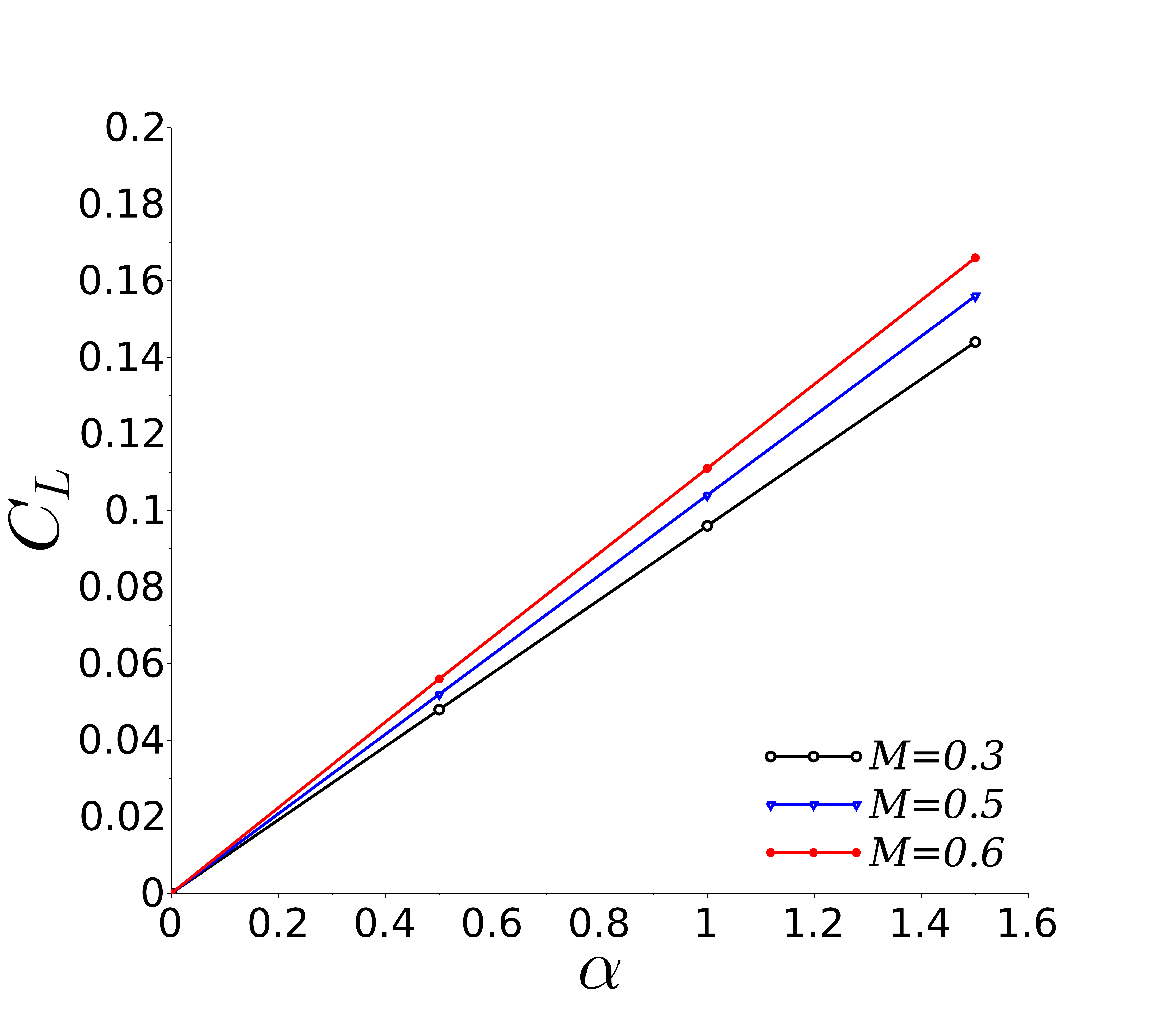}
    \includegraphics[width=0.49\textwidth]{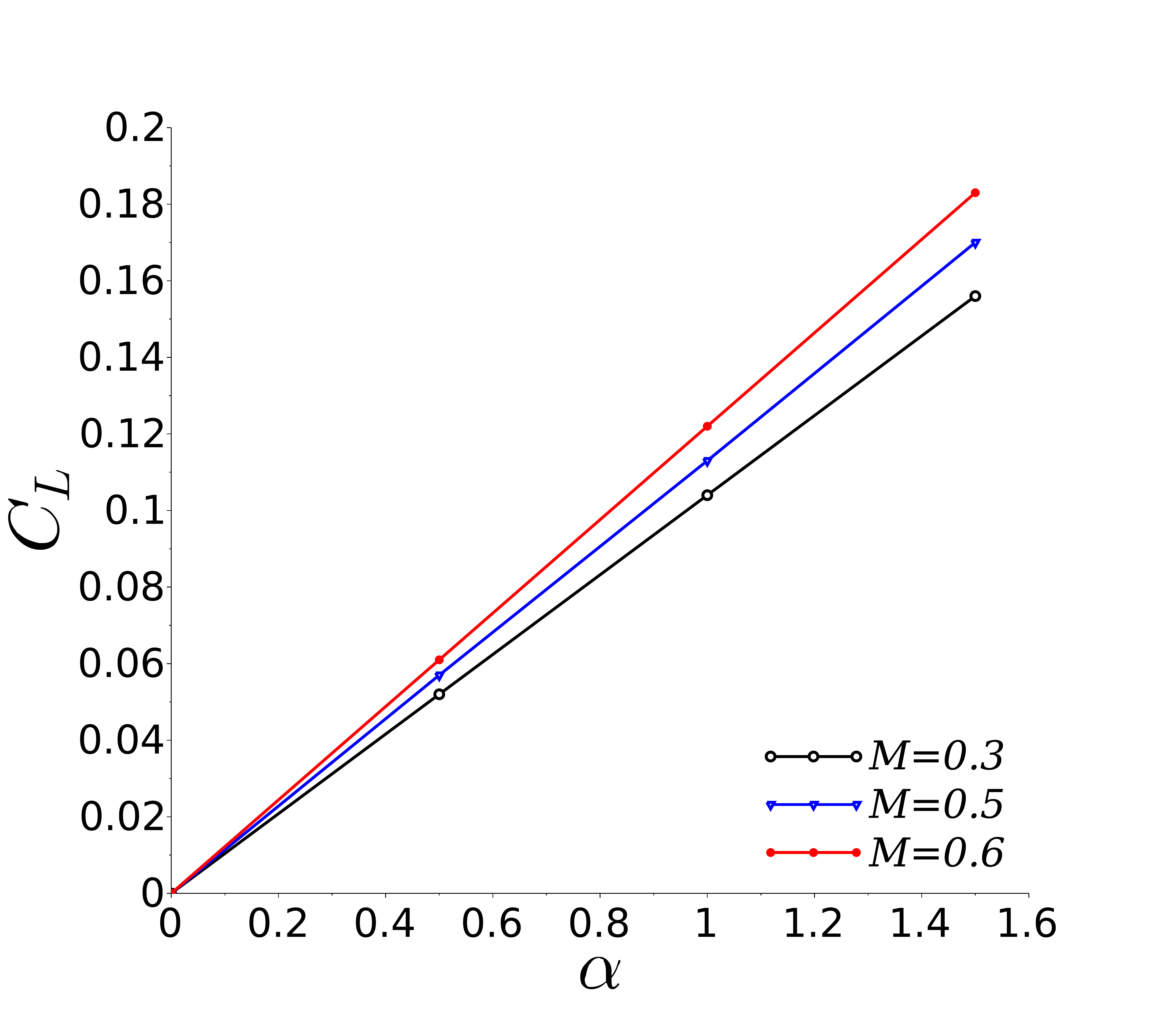}
    \caption{Results of steady-state simulations. Upper row: $\eta=6$ (left) and $\eta=8$ (right); lower row: $\eta=12$ (left) and $\eta=20$ (right).}
    \label{fig:cla}
\end{figure}

First, steady-state simulations have been performed for $0<\alpha<1.5$ and results are shown in Figure \ref{fig:cla}; indeed, their linearity is fully consistent with that of the time-accurate simulations shown in Figure \ref{fig:convergence}. As expected for all Mach numbers, the lift derivative of an isolated two-dimensional section is gradually approached with increasing the wing aspect ratio, due to the decreasing downwash.

\begin{figure}
    \centering
    \includegraphics[width=0.49\textwidth]{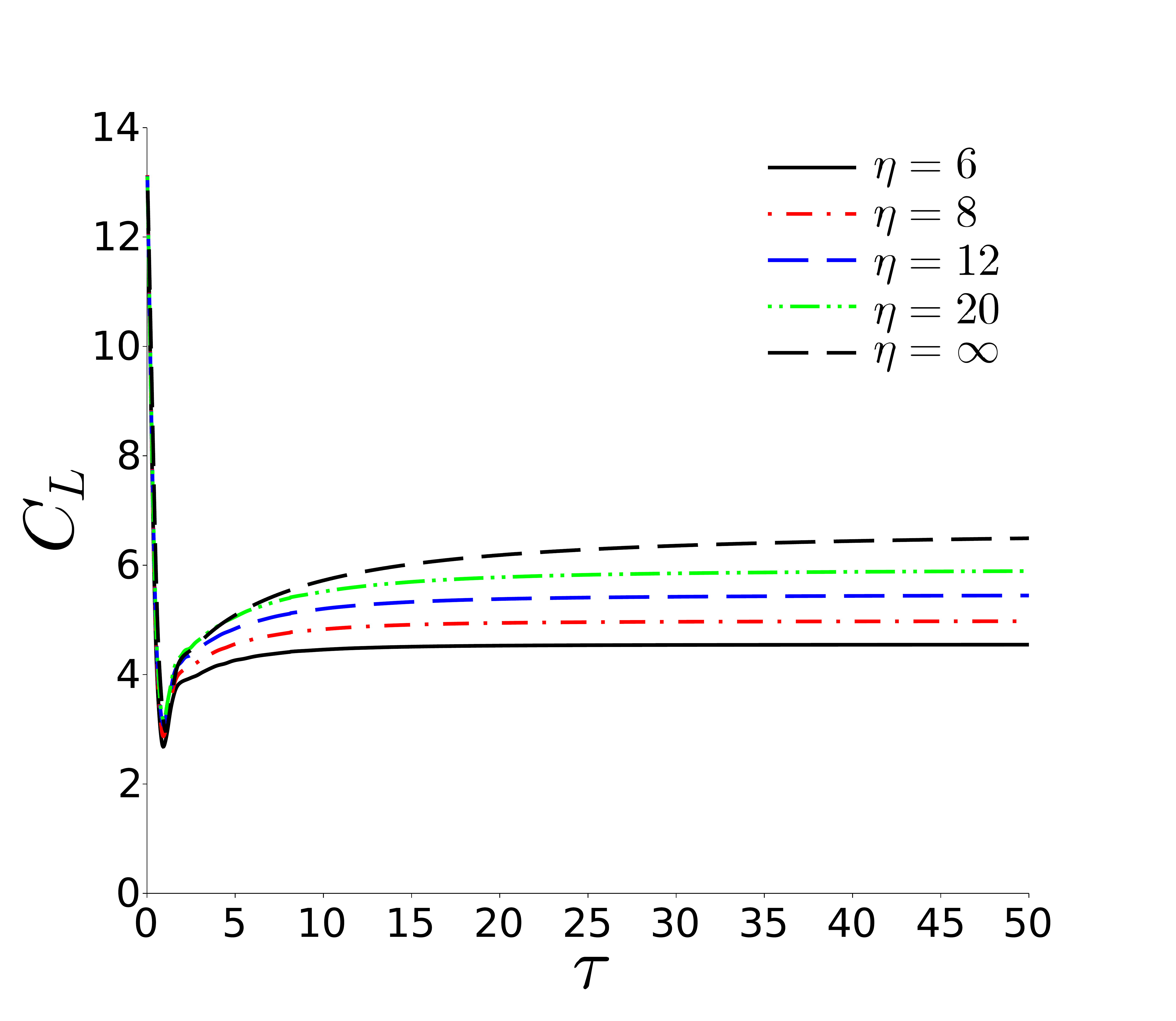}
    \includegraphics[width=0.49\textwidth]{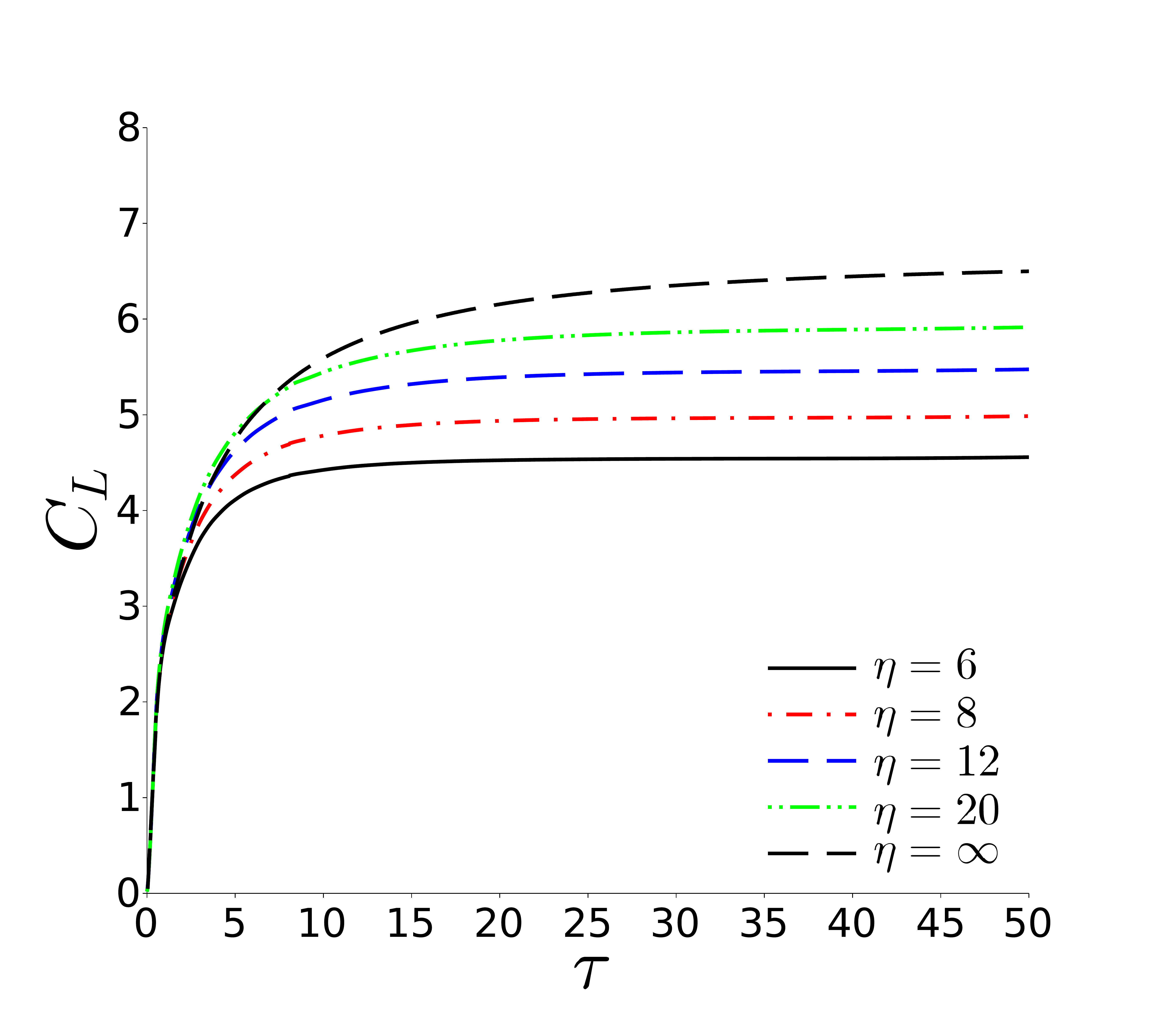}
    \caption{Wing lift coefficient for unit AoA step (left) and unit SEG (right) for M=0.3.}
    \label{fig:cfd_allars_m03}
\end{figure}

\begin{figure}
    \centering
    \includegraphics[width=0.49\textwidth]{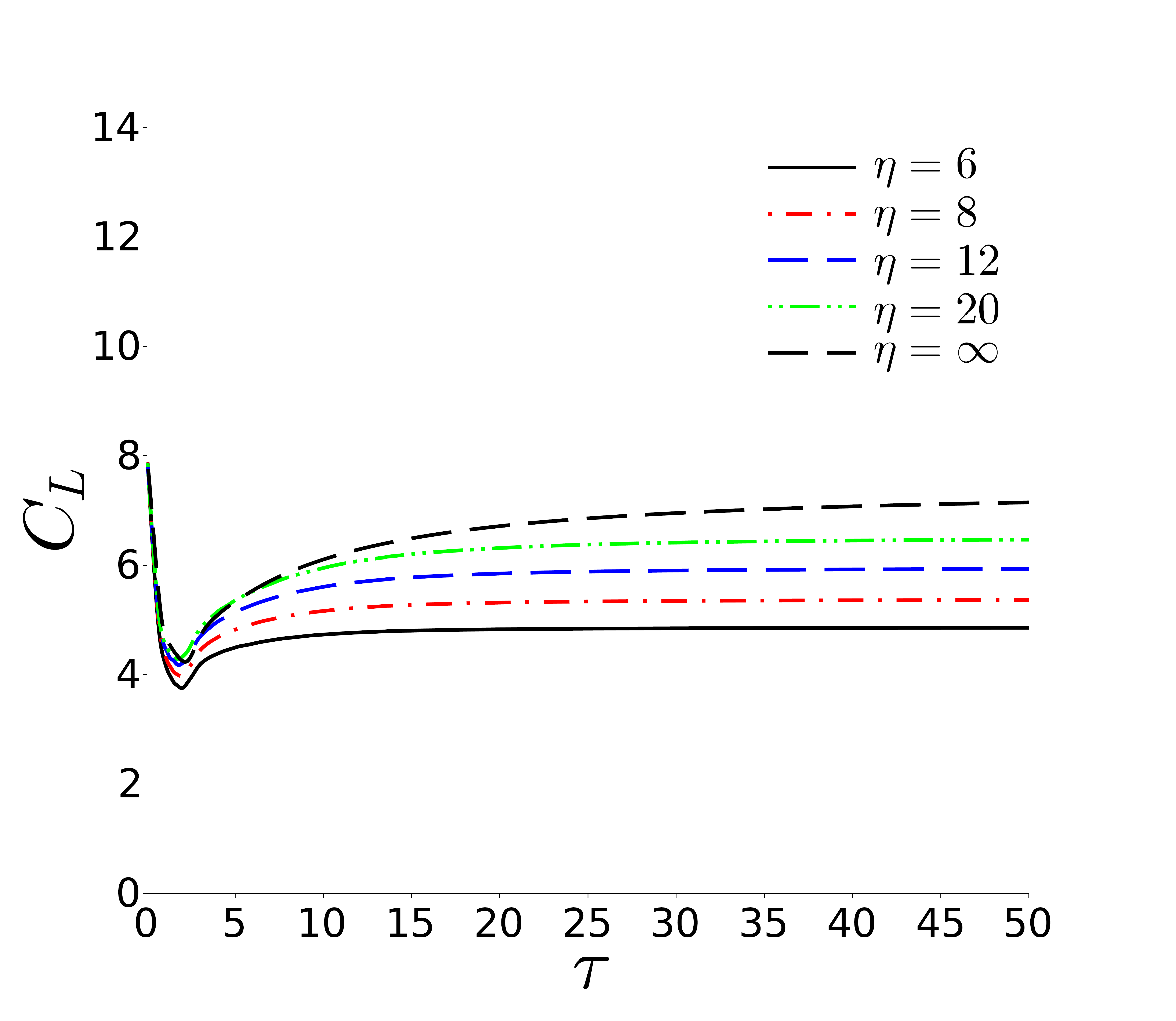}
    \includegraphics[width=0.49\textwidth]{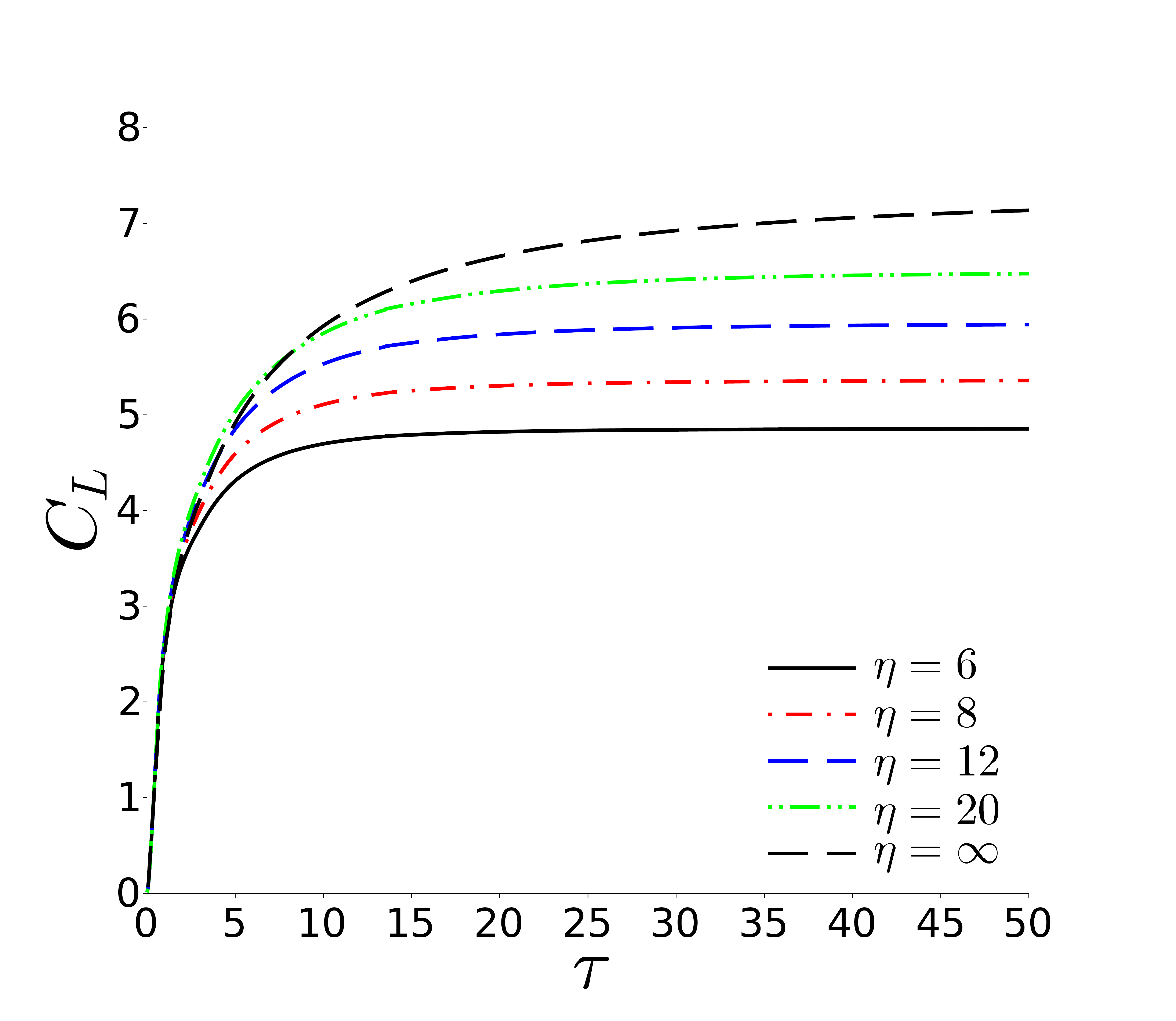}
    \caption{Wing lift coefficient for unit AoA step (left) and unit SEG (right) for M=0.5.}
    \label{fig:cfd_allars_m05}
\end{figure}

\begin{figure}
    \centering
    \includegraphics[width=0.49\textwidth]{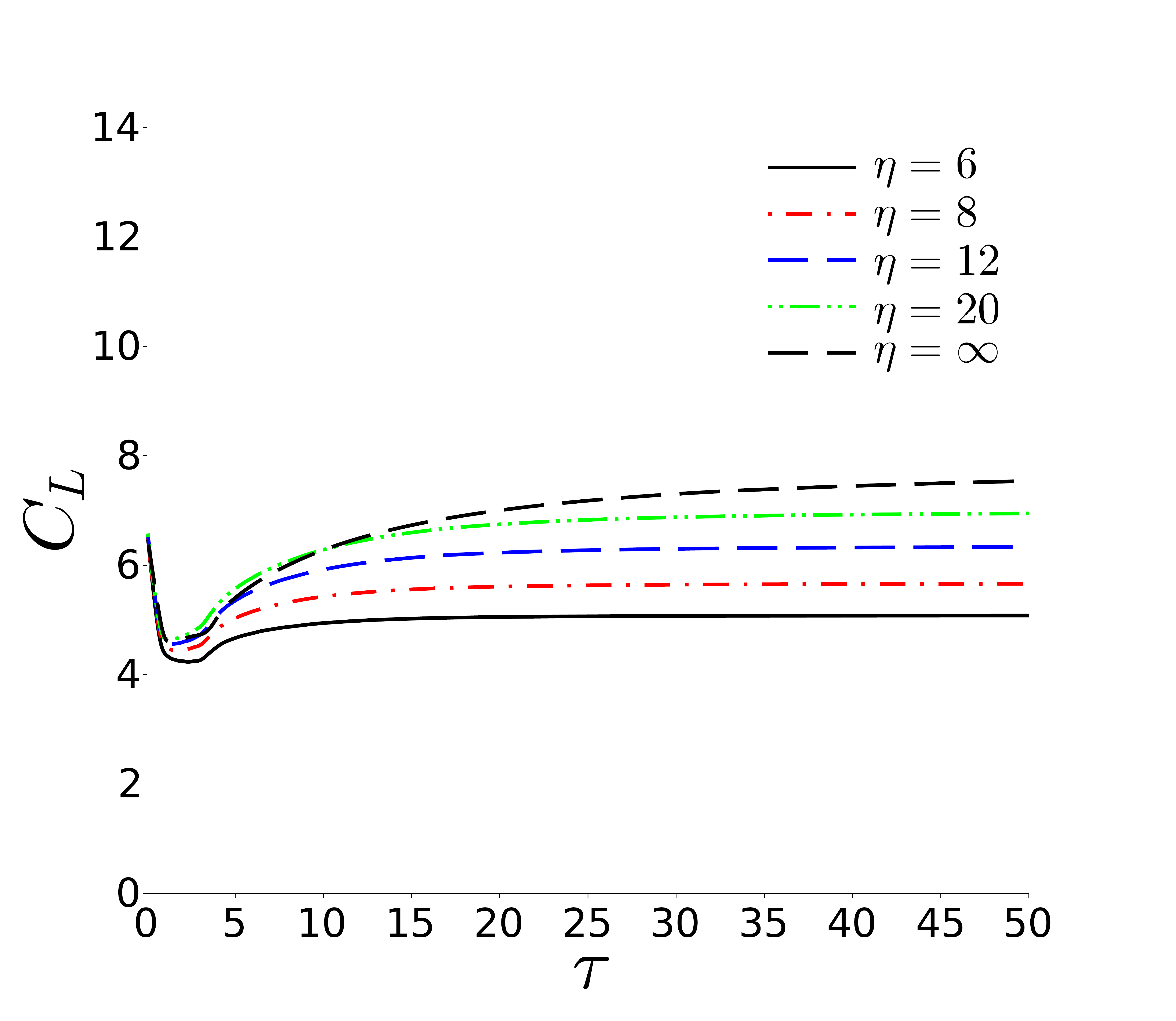}
    \includegraphics[width=0.49\textwidth]{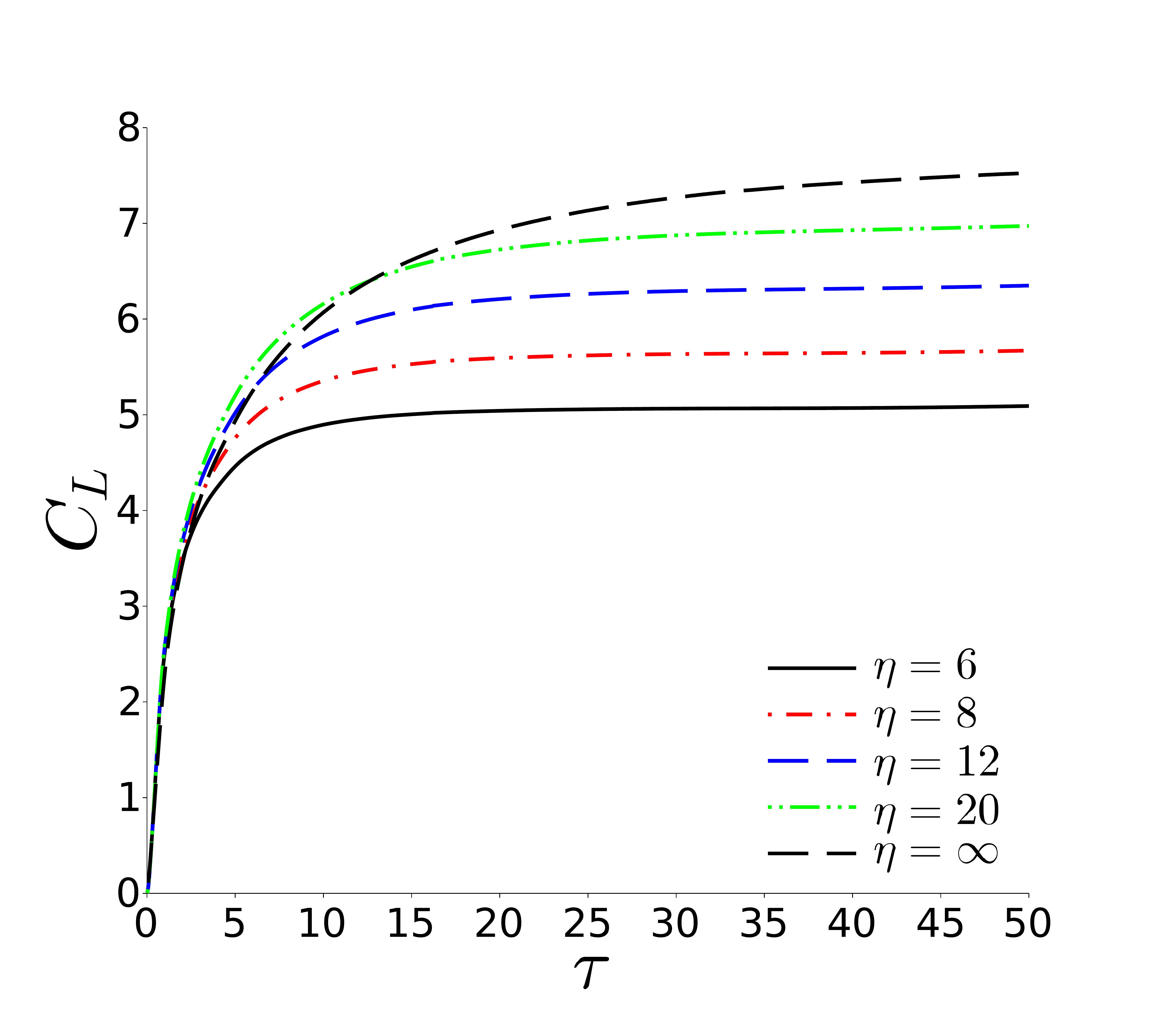}
    \caption{Wing lift coefficient for unit AoA step (left) and unit SEG (right) for M=0.6.}
    \label{fig:cfd_allars_m06}
\end{figure}

Then, unsteady simulations have been performed and results are shown in Figures \ref{fig:cfd_allars_m03}, \ref{fig:cfd_allars_m05} and \ref{fig:cfd_allars_m06}, for both a unit step in the angle of attack and a vertical sharp-edge gust. Still for all Mach numbers, the lift build-up of an isolated two-dimensional section is gradually approached with increasing the wing aspect ratio; yet, note that most of the downwash effect lays in the circulatory part of the flow response where the two-dimensional lift build-up takes longer than that of the elliptical wings (reaching their asymptotic condition quite earlier, even with $\eta=20$), for which the non-circulatory part decays more rapidly due to geometric taper.

\subsection{Tuning of the parametric analytical approximation}
\label{subsec:tuning}

The analytical approximations have been tuned for rigorous comparisons with the CFD results, within a consistent framework where initial rate and final value of the lift development are extracted from the latter.

\begin{figure}
    \centering
    \includegraphics[width=0.49\textwidth]{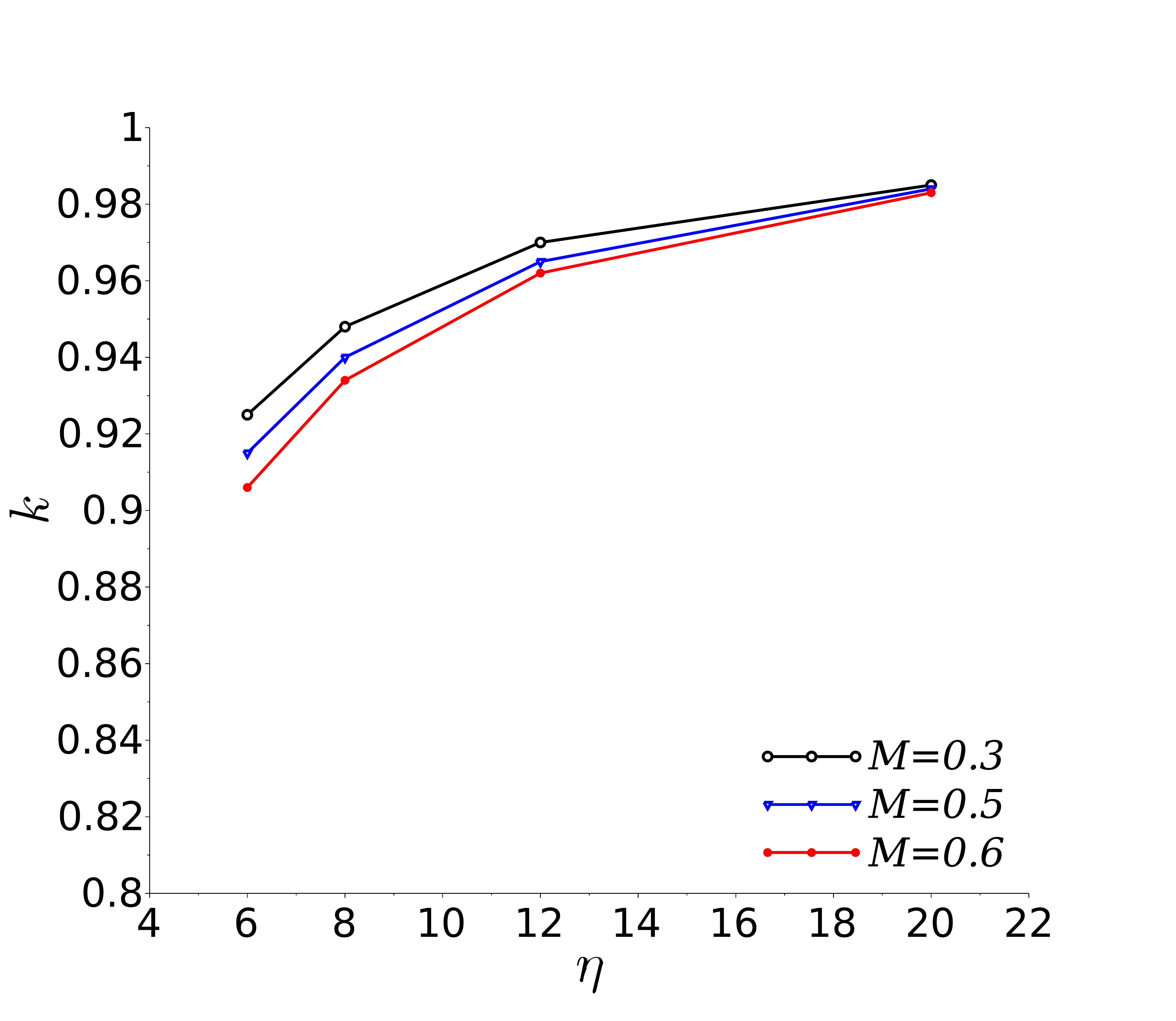}
    \includegraphics[width=0.49\textwidth]{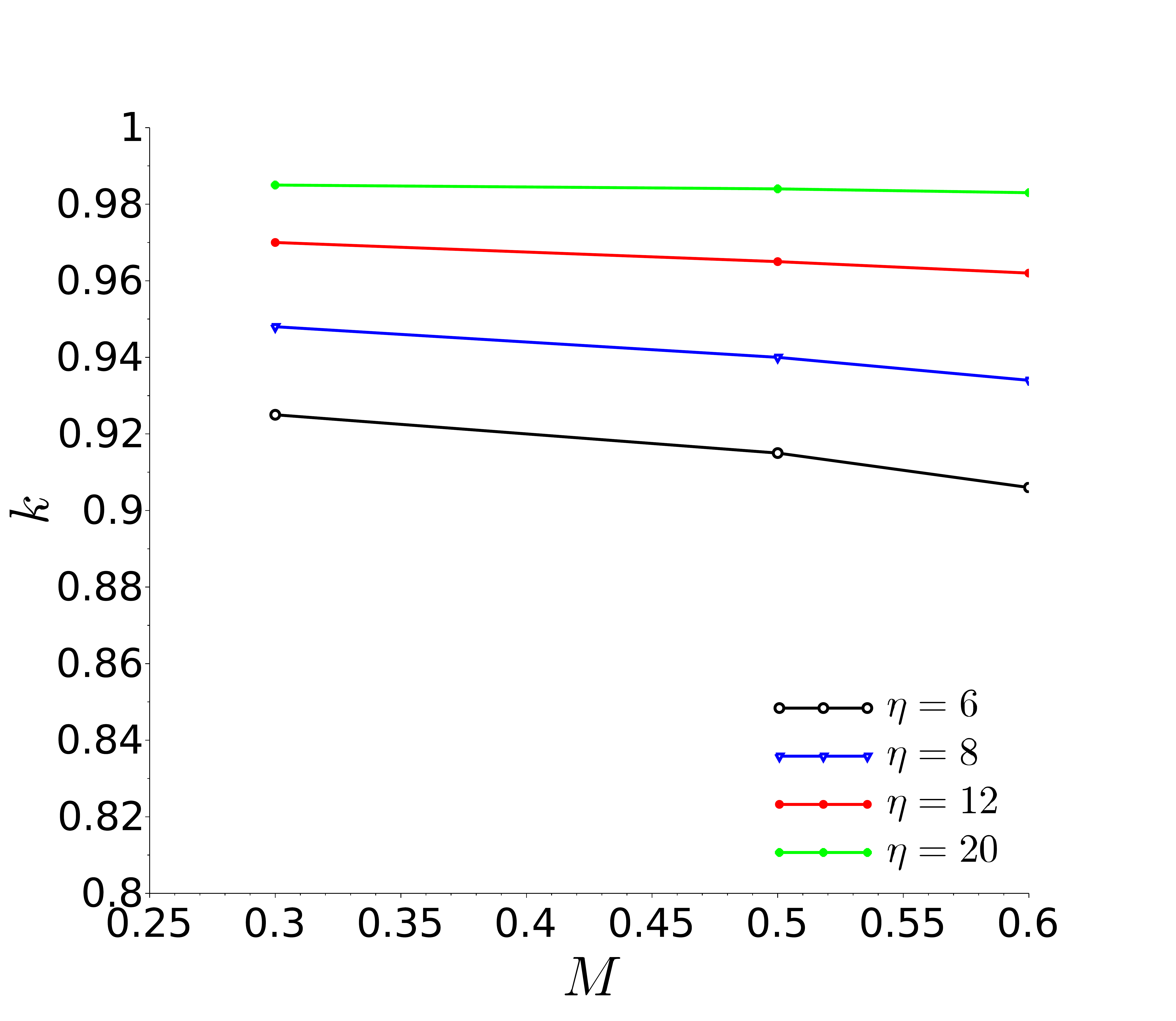}
    \caption{Tuning factor for the final steady lift $\bar{C}_L$, per Mach number (left) and aspect ratio (right).}
    \label{fig:coeff_tab_ksteady}
\end{figure}

The steady slope of the lift curve has been assessed by CFD for each aspect ratio and Mach number. As expected, the numerical results show slightly smaller values than the theoretical ones (which include aerofoil thickness effects but disregard the stronger downwash at the wing tips); the ratios are shown in Figure \ref{fig:coeff_tab_ksteady} and tend to unity with increasing aspect ratio and decreasing Mach number.

\begin{figure}
    \centering
    \includegraphics[width=0.49\textwidth]{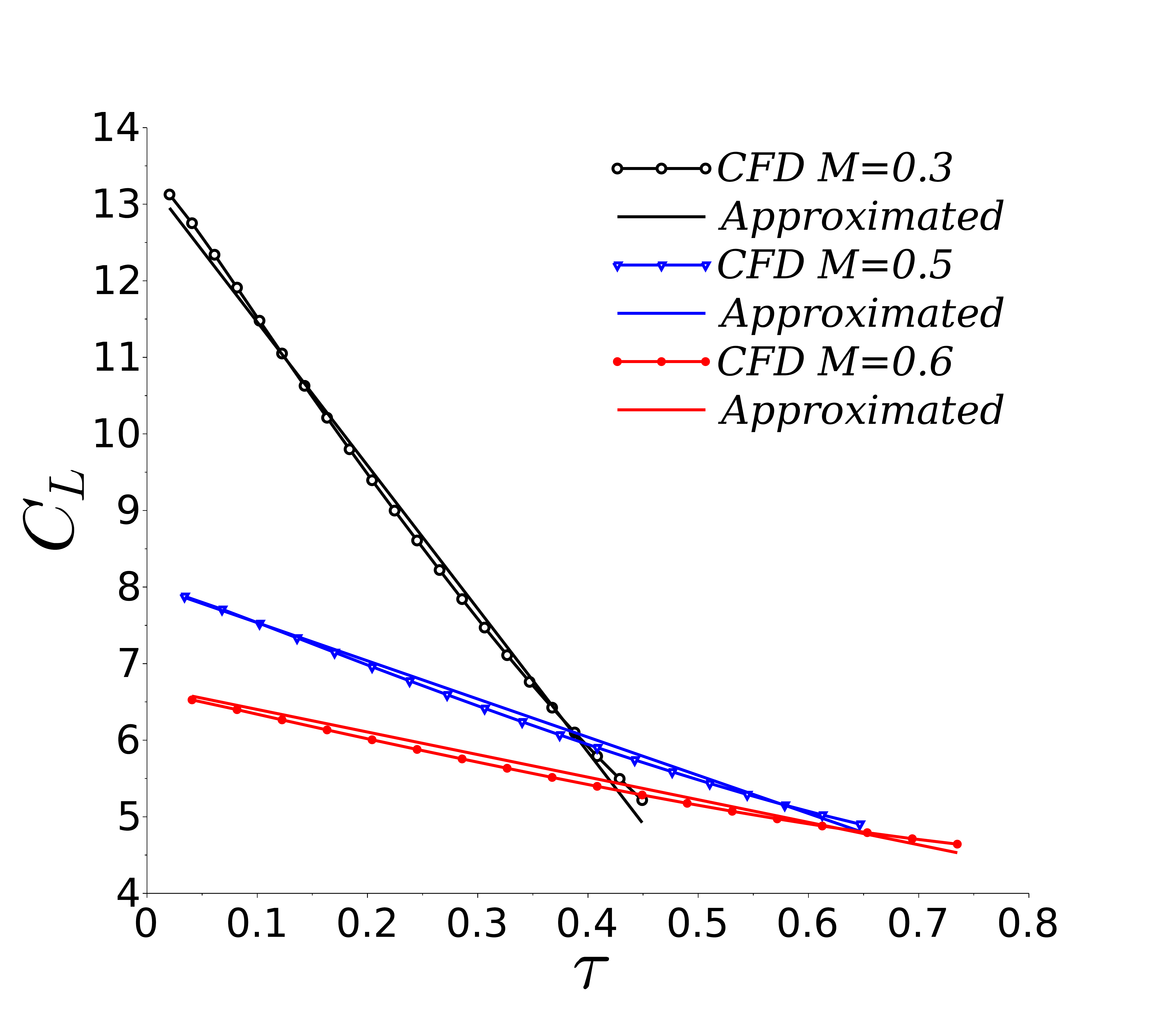}
    \includegraphics[width=0.49\textwidth]{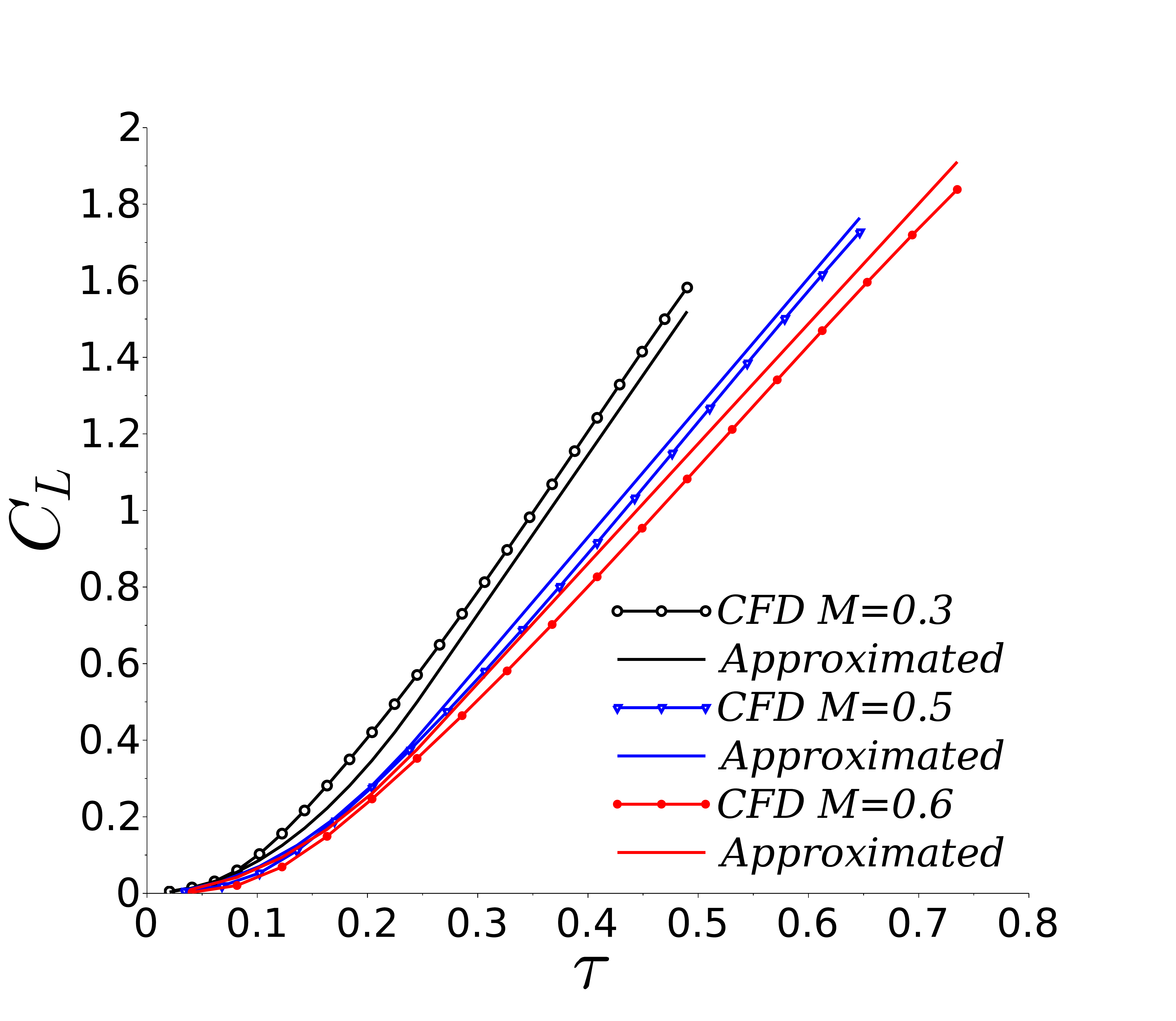}
    \caption{Assessment of the initial lift rate for $\eta=6$, unit AoA step (left) and unit SEG (right)}
    \label{fig:initial_ar6}
\end{figure}

\begin{figure}
    \centering
    \includegraphics[width=0.49\textwidth]{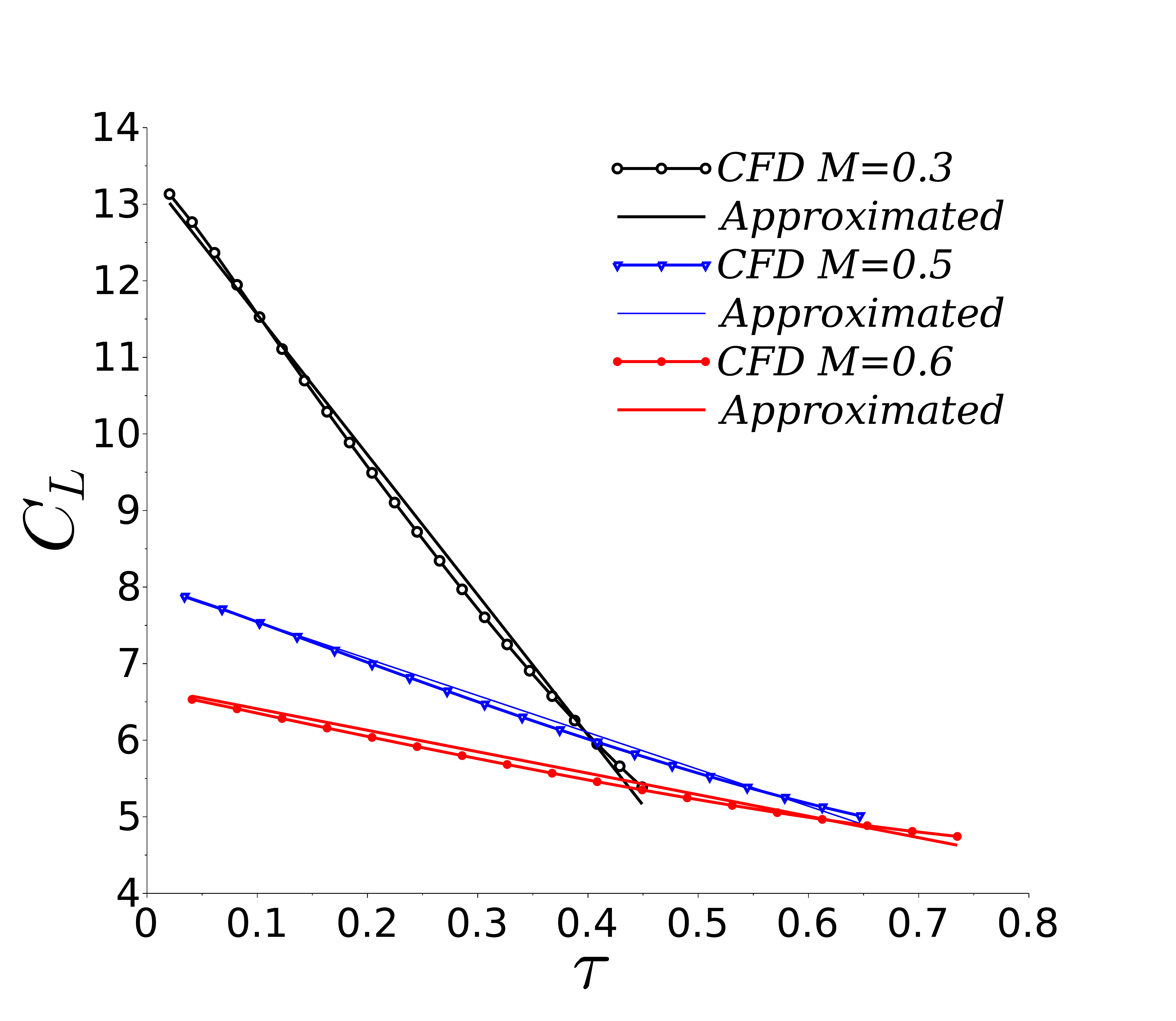}
    \includegraphics[width=0.49\textwidth]{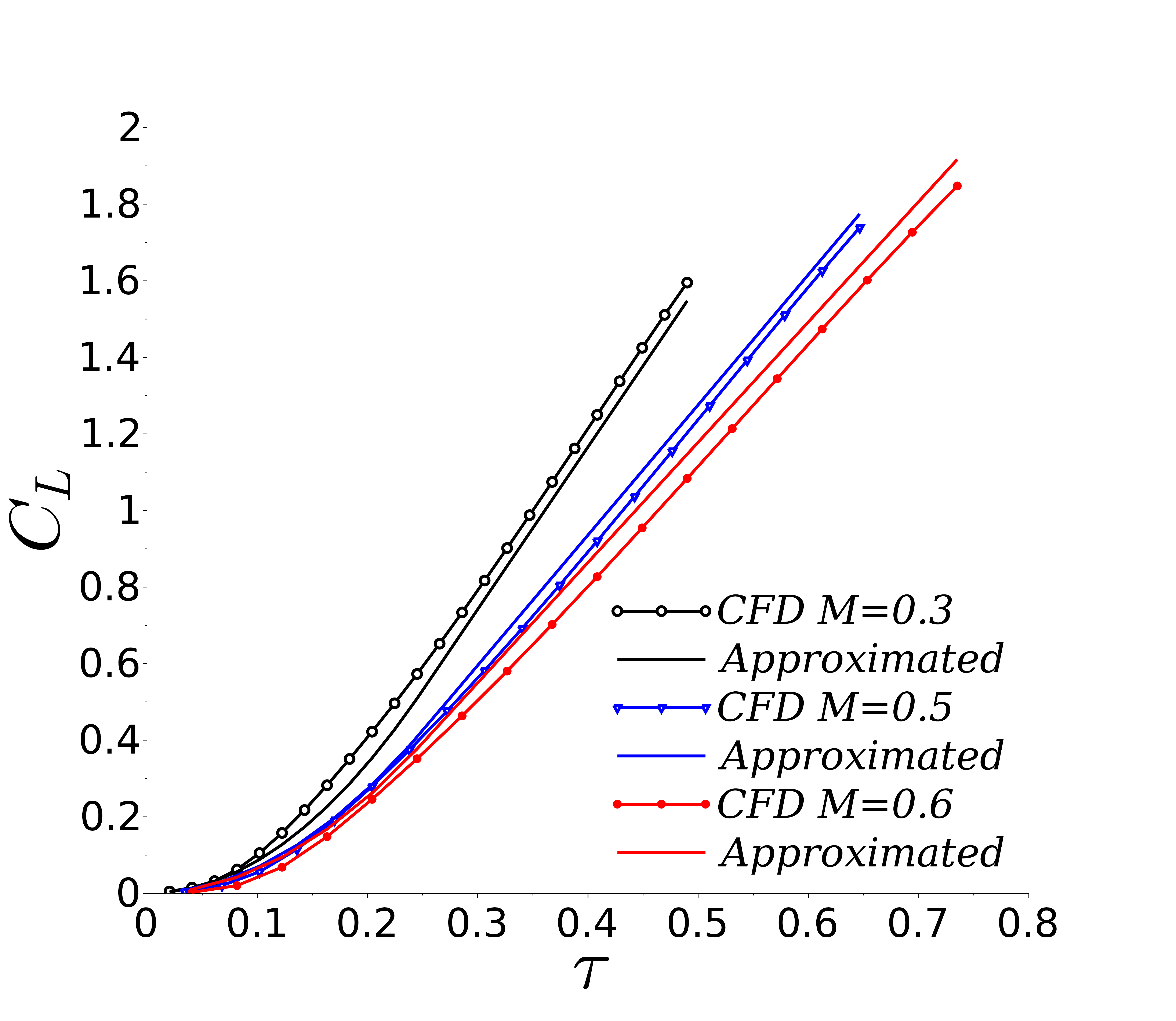}
    \caption{Assessment of the initial lift rate for $\eta=8$, unit AoA step (left) and unit SEG (right)}
    \label{fig:initial_ar8}
\end{figure}

\begin{figure}
    \centering
    \includegraphics[width=0.49\textwidth]{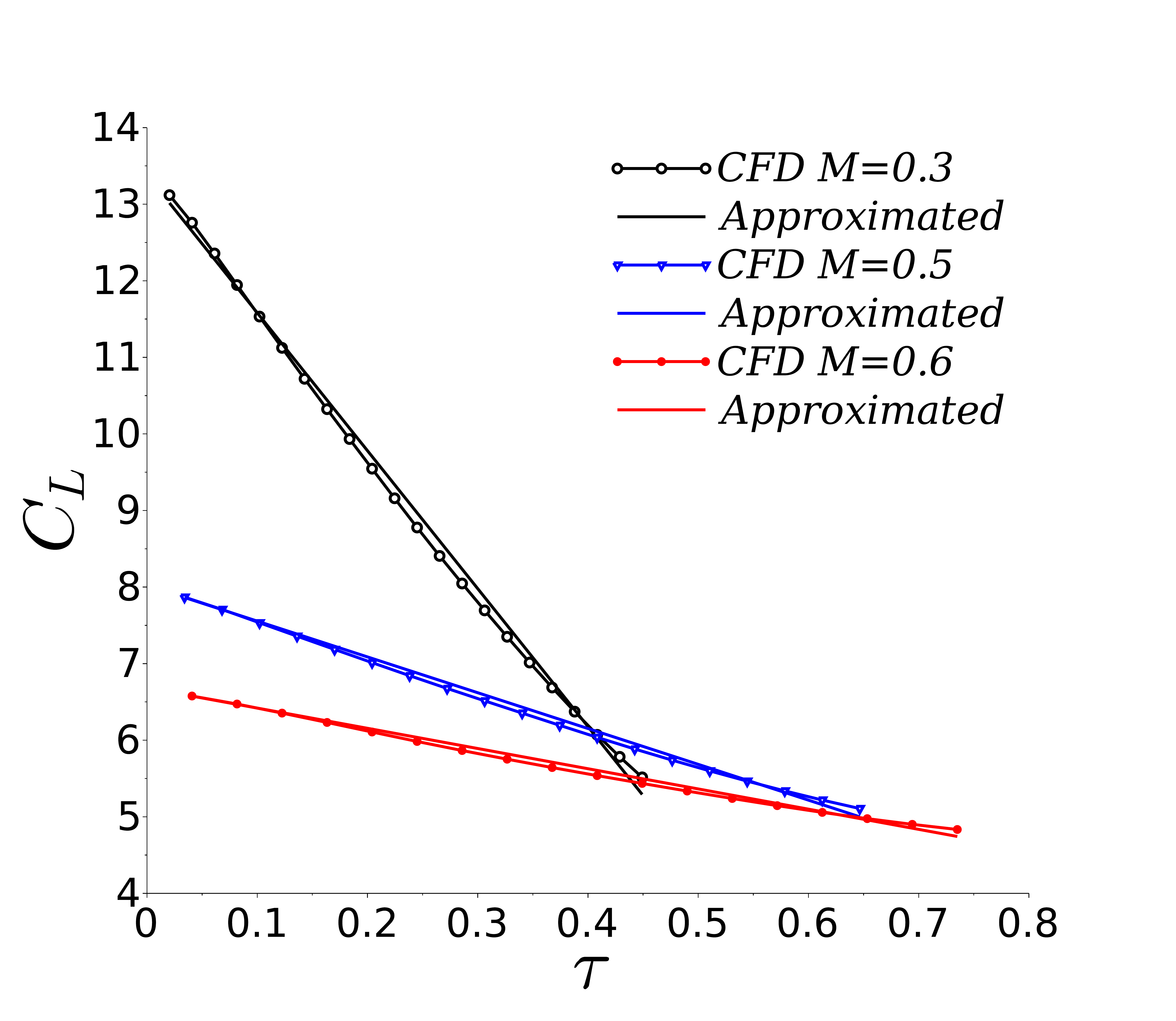}
    \includegraphics[width=0.49\textwidth]{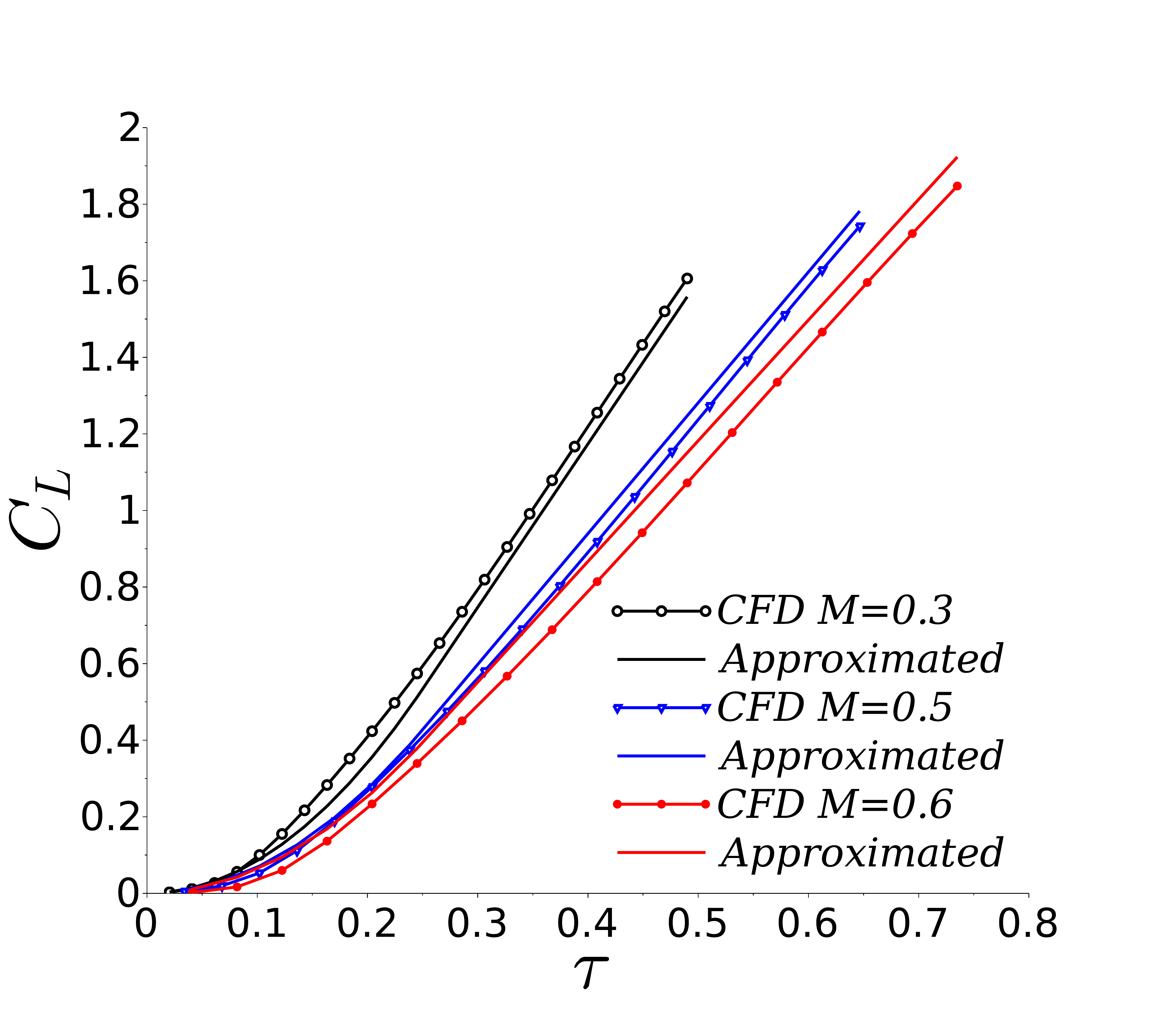}
    \caption{Assessment of the initial lift rate for $\eta=12$, unit AoA step (left) and unit SEG (right)}
    \label{fig:initial_ar12}
\end{figure}

\begin{figure}
    \centering
    \includegraphics[width=0.49\textwidth]{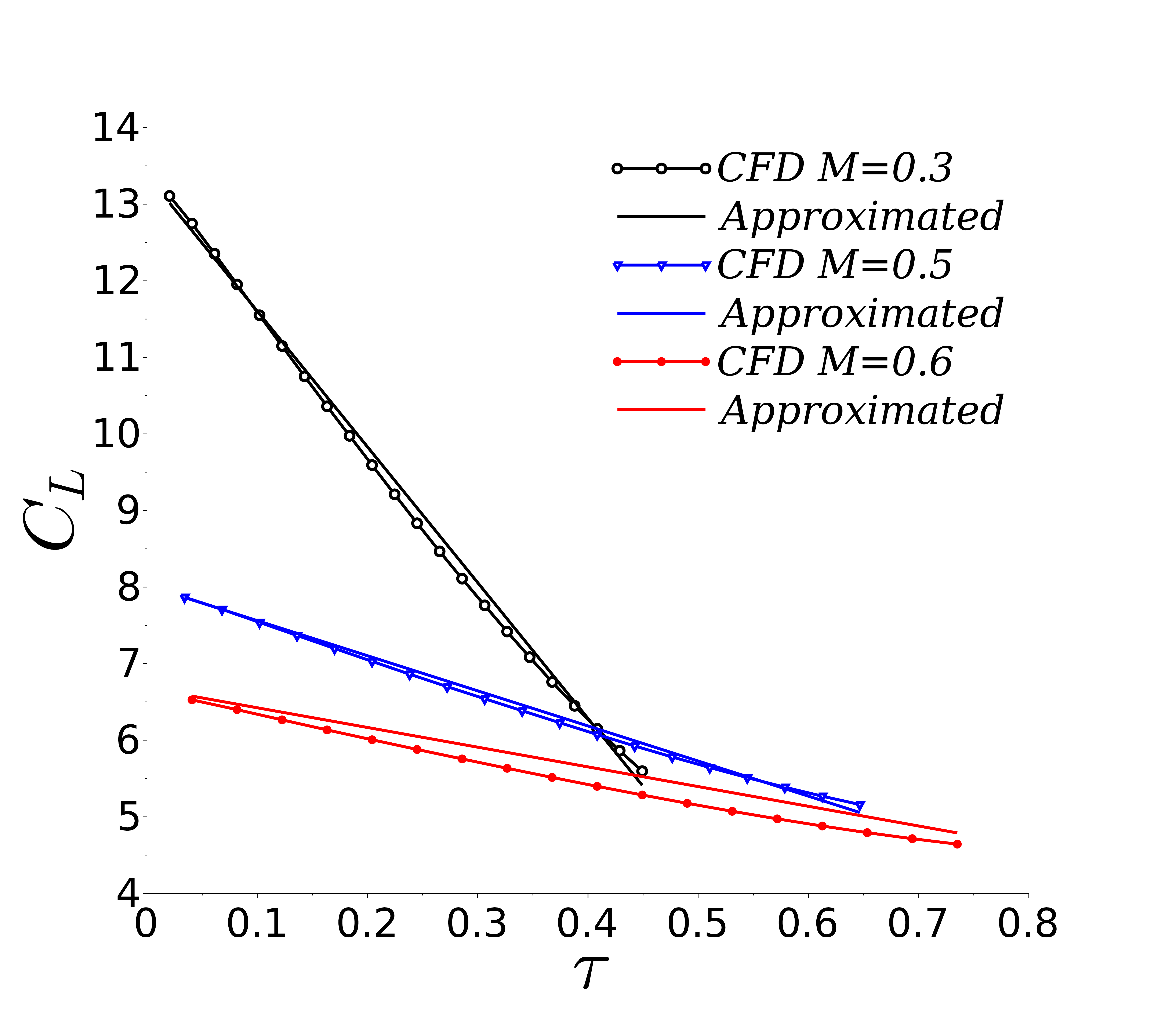}
    \includegraphics[width=0.49\textwidth]{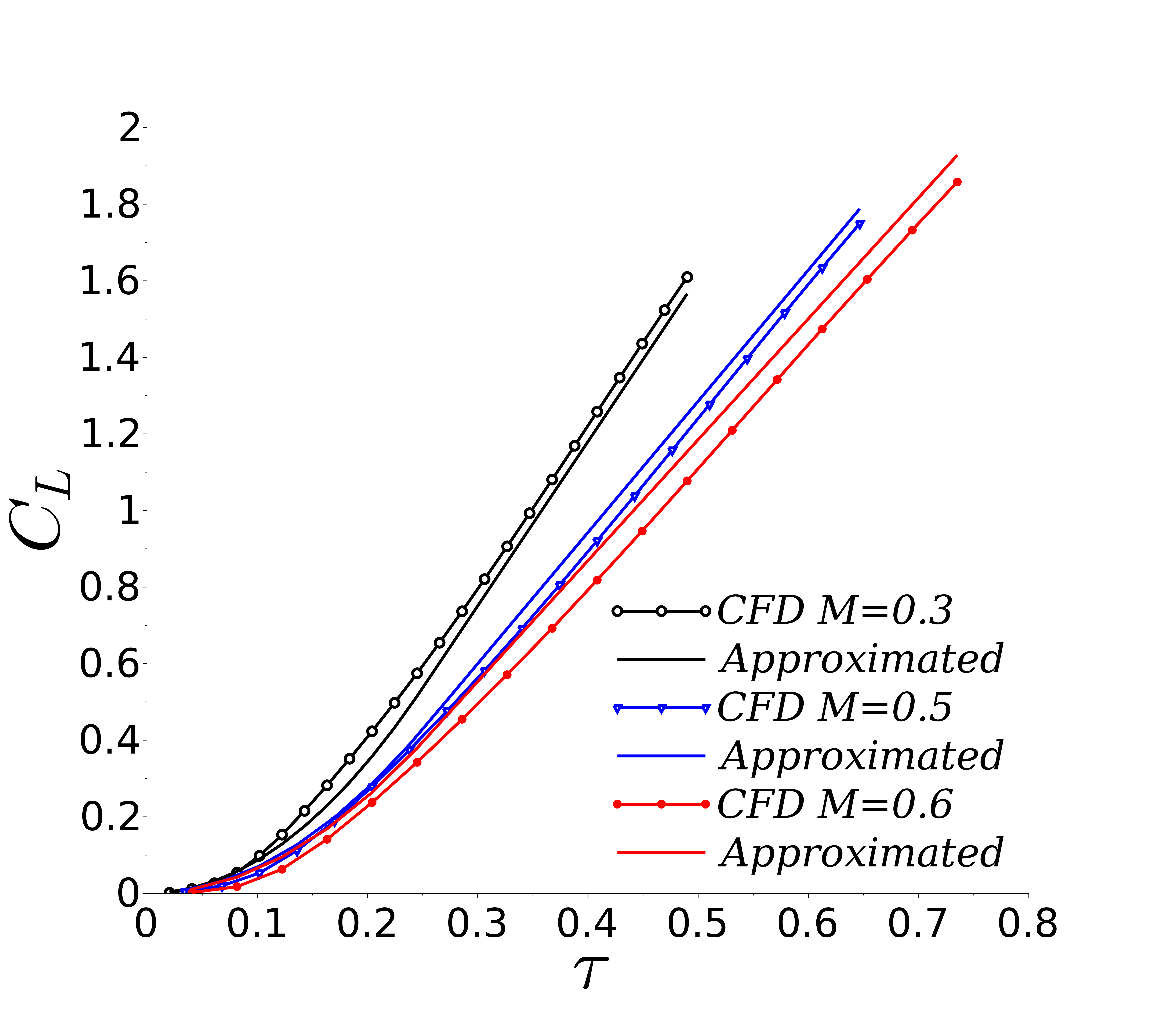}
    \caption{Assessment of the initial lift rate for $\eta=20$, unit AoA step (left) and unit SEG (right)}
    \label{fig:initial_ar20}
\end{figure}

\begin{figure}
    \centering
    \includegraphics[width=0.49\textwidth]{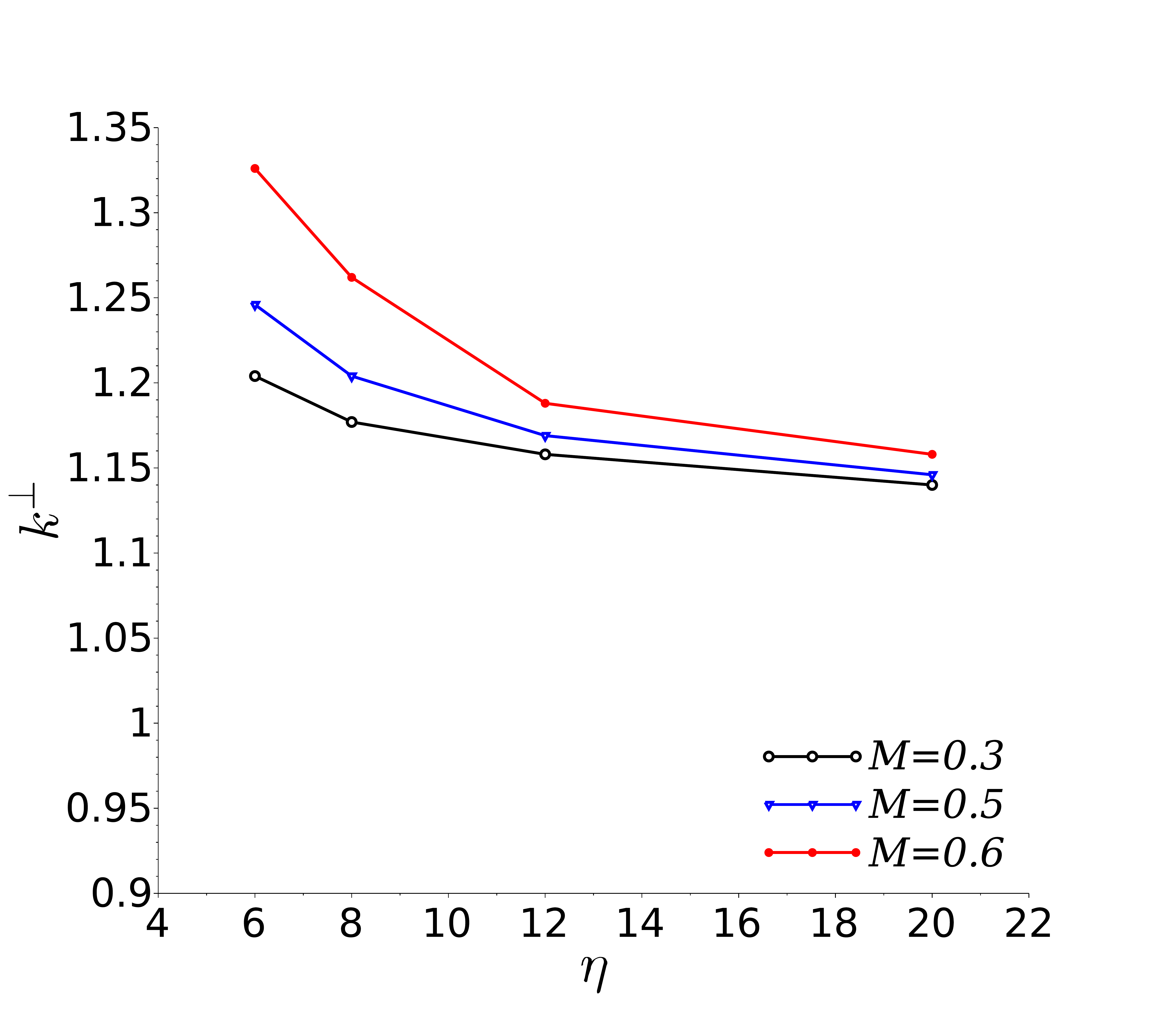}
    \includegraphics[width=0.49\textwidth]{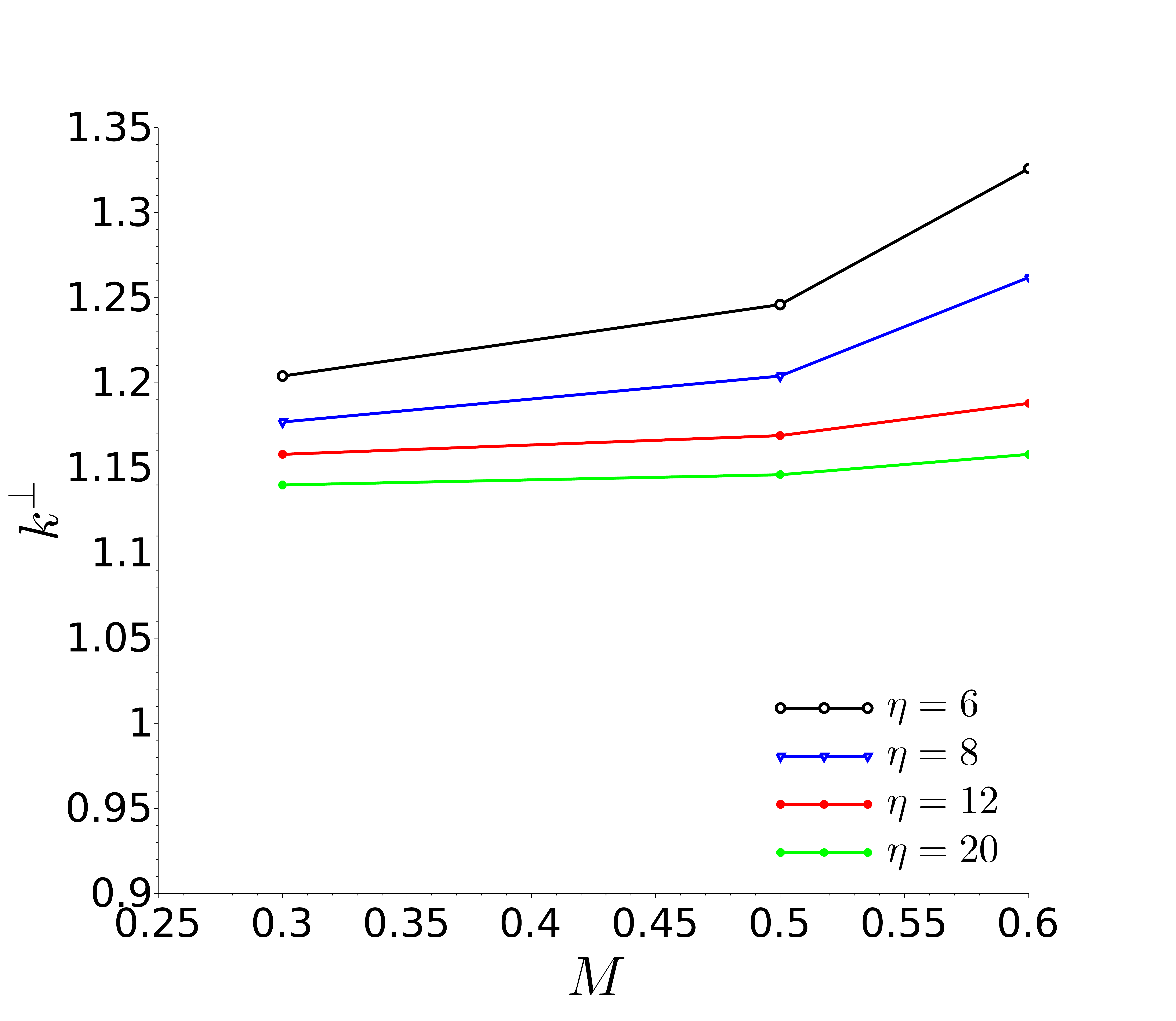}
    \caption{Tuning factors for the initial lift rate due to a unit AoA step, per Mach number (left) and aspect ratio (right).}
    \label{fig:coeff_tab_aoa}
\end{figure}

\begin{figure}
    \centering
    \includegraphics[width=0.49\textwidth]{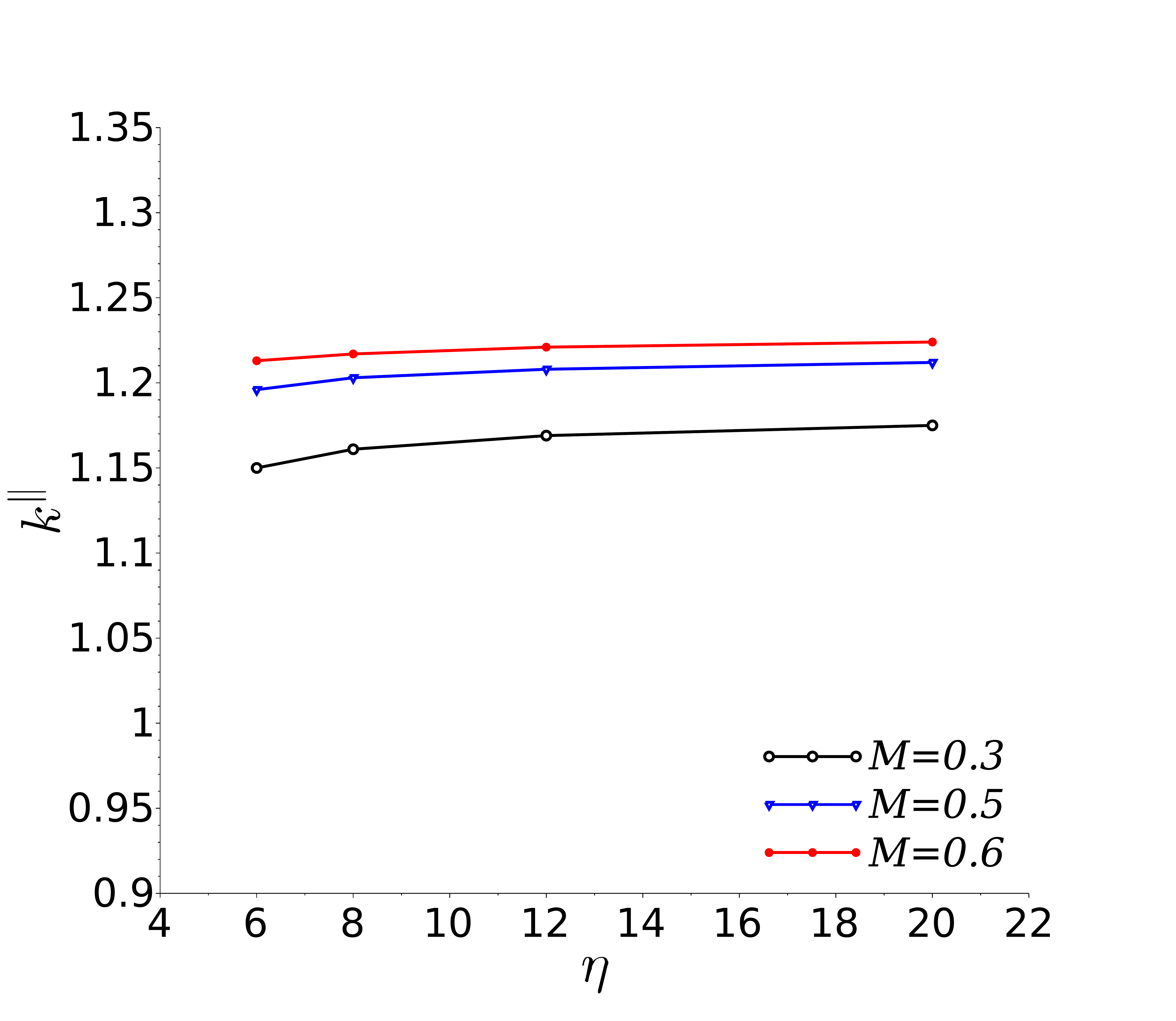}
    \includegraphics[width=0.49\textwidth]{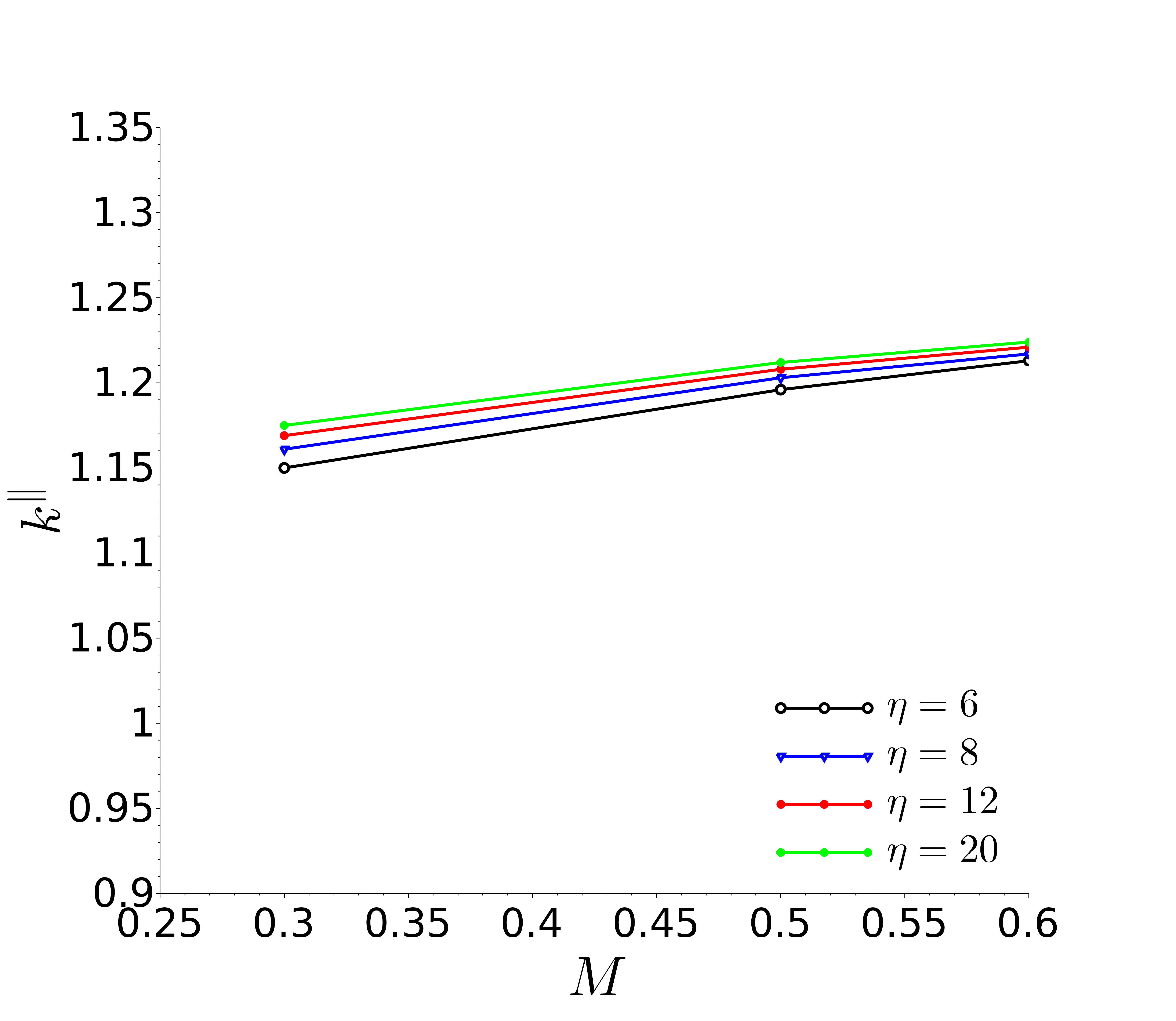}
    \caption{Tuning factors for the initial lift rate due to a unit SEG, per Mach number (left) and aspect ratio (right).}
    \label{fig:coeff_tab_seg}
\end{figure}

The initial lift rate in the impulsive part of the flow response (to both flow perturbations) for a three dimensional wing is different from the two-dimensional one, although the starting lift value is identical. This is because the evolution of the pressure waves is three-dimensional (hence depending on wing geometry), whereas the starting lift value only depends on the flow perturbation magnitude, which is two-dimensional. The CFD solutions show that the lift of elliptical wings decays more rapidly than that of their isolated aerofoil; the ratio of the initial slopes is shown in Figures \ref{fig:coeff_tab_aoa} and \ref{fig:coeff_tab_seg} for the unit AoA and SEG perturbations, respectively. The derivation of the tuning coefficients has been carried out on the basis of a conventional linear regression; original CFD curves and tuned analytical solutions are shown in Figures \ref{fig:initial_ar6} to \ref{fig:initial_ar20}. Note that the simultaneous chord- and span-wise gust penetration has indeed been taken into account in the correct derivation of the initial lift rate.


\subsection{Comparisons}
\label{subsec:responses}

\begin{figure}
\captionsetup{justification=raggedright,singlelinecheck=false}
\begin{center}
\includegraphics[width=0.49\textwidth]{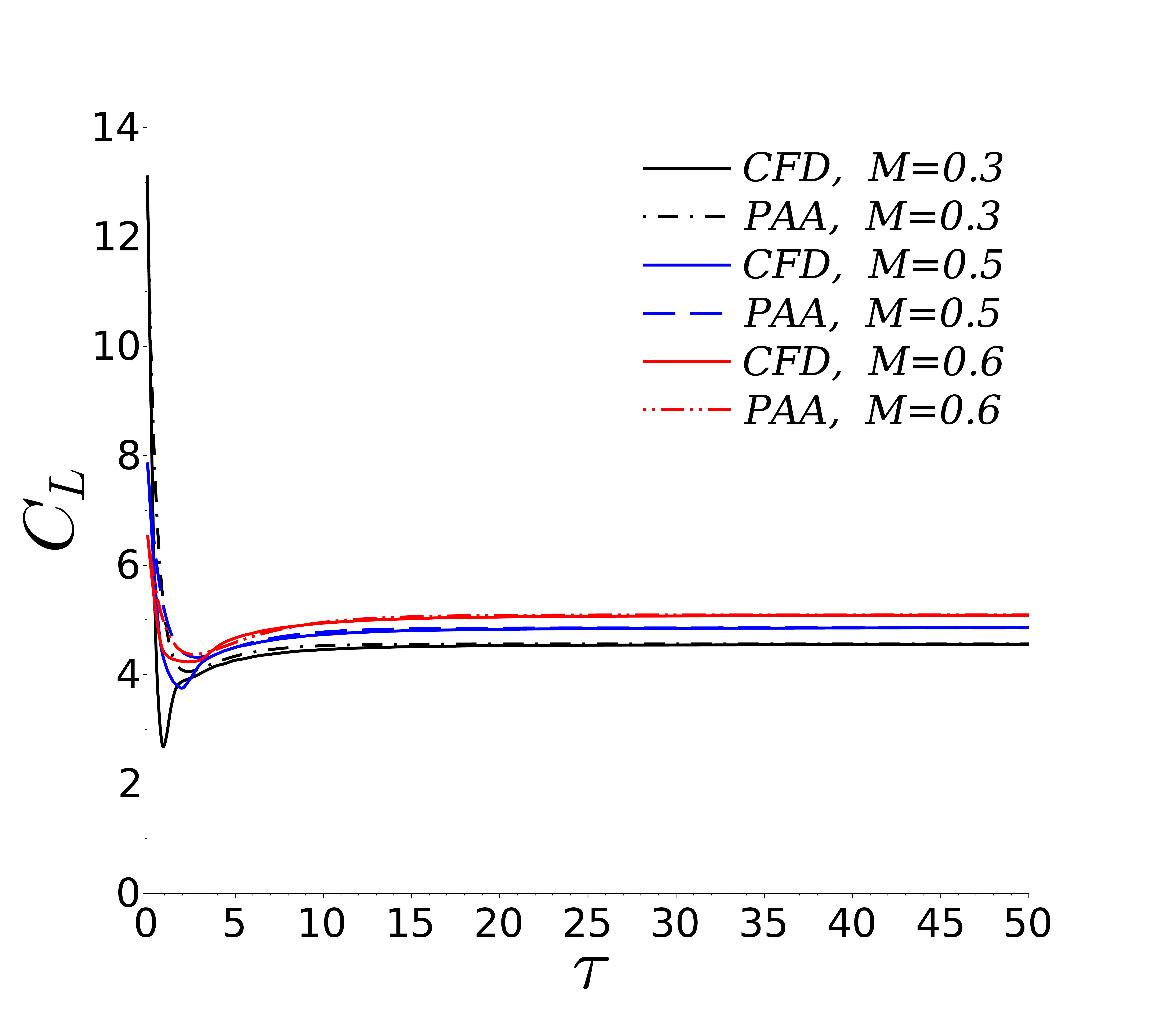}
\includegraphics[width=0.49\textwidth]{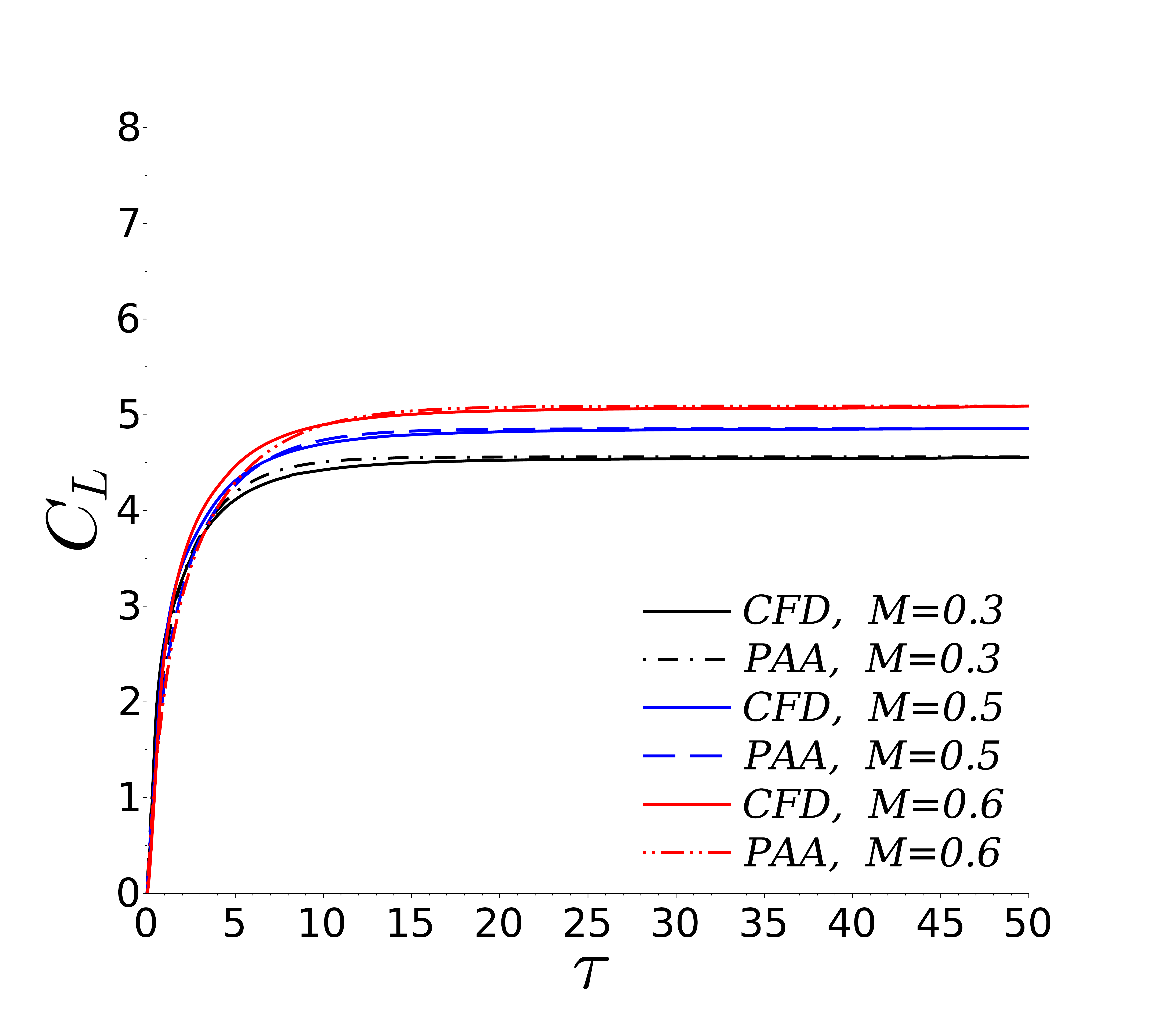}
\end{center}
\caption{Wing lift coefficient for $\eta=6$, unit AoA step (left) and unit SEG (right).}
\label{fig:aoa_ar6}
\end{figure}

\begin{figure}
\begin{center}
\includegraphics[width=0.49\textwidth]{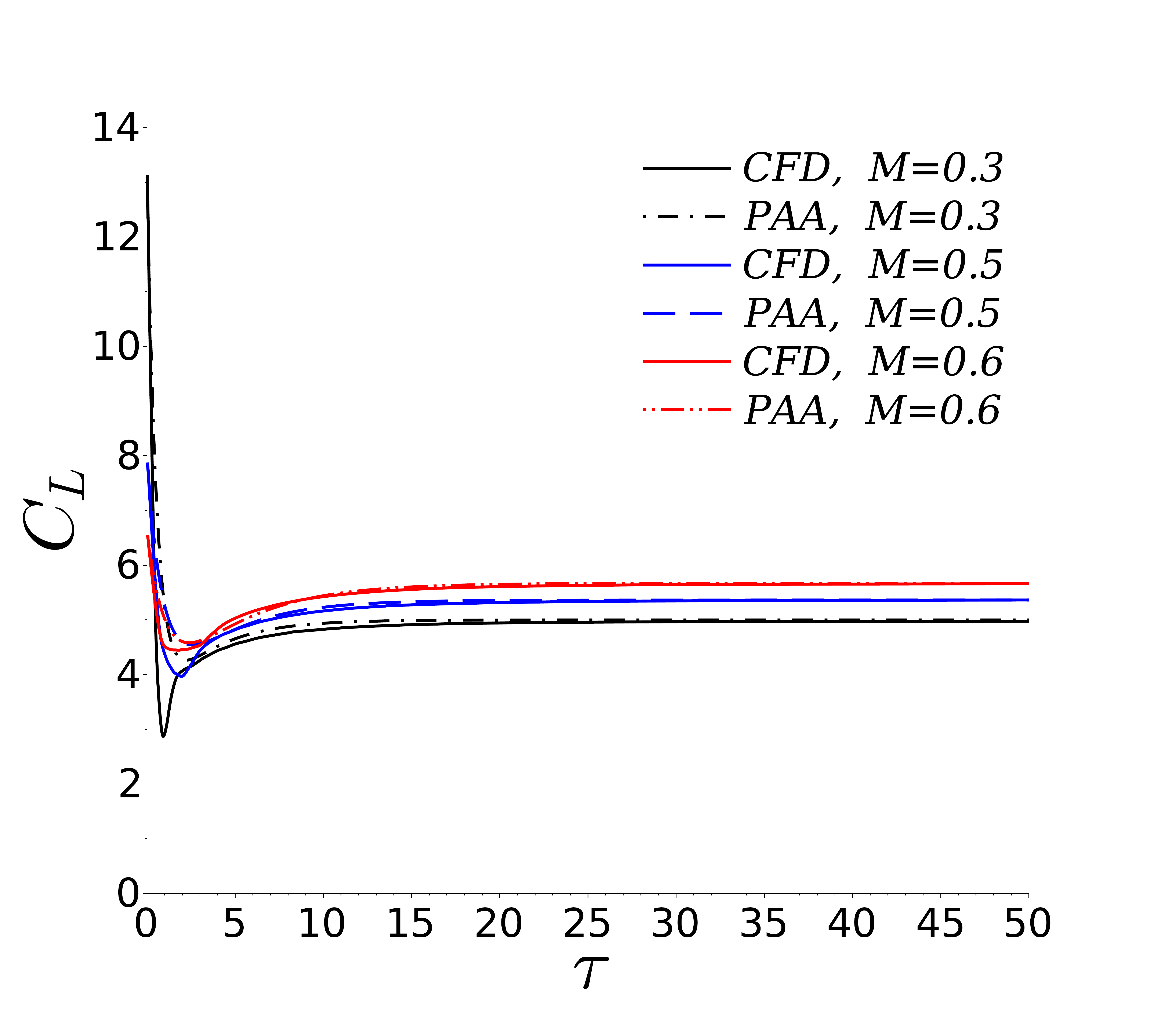}
\includegraphics[width=0.49\textwidth]{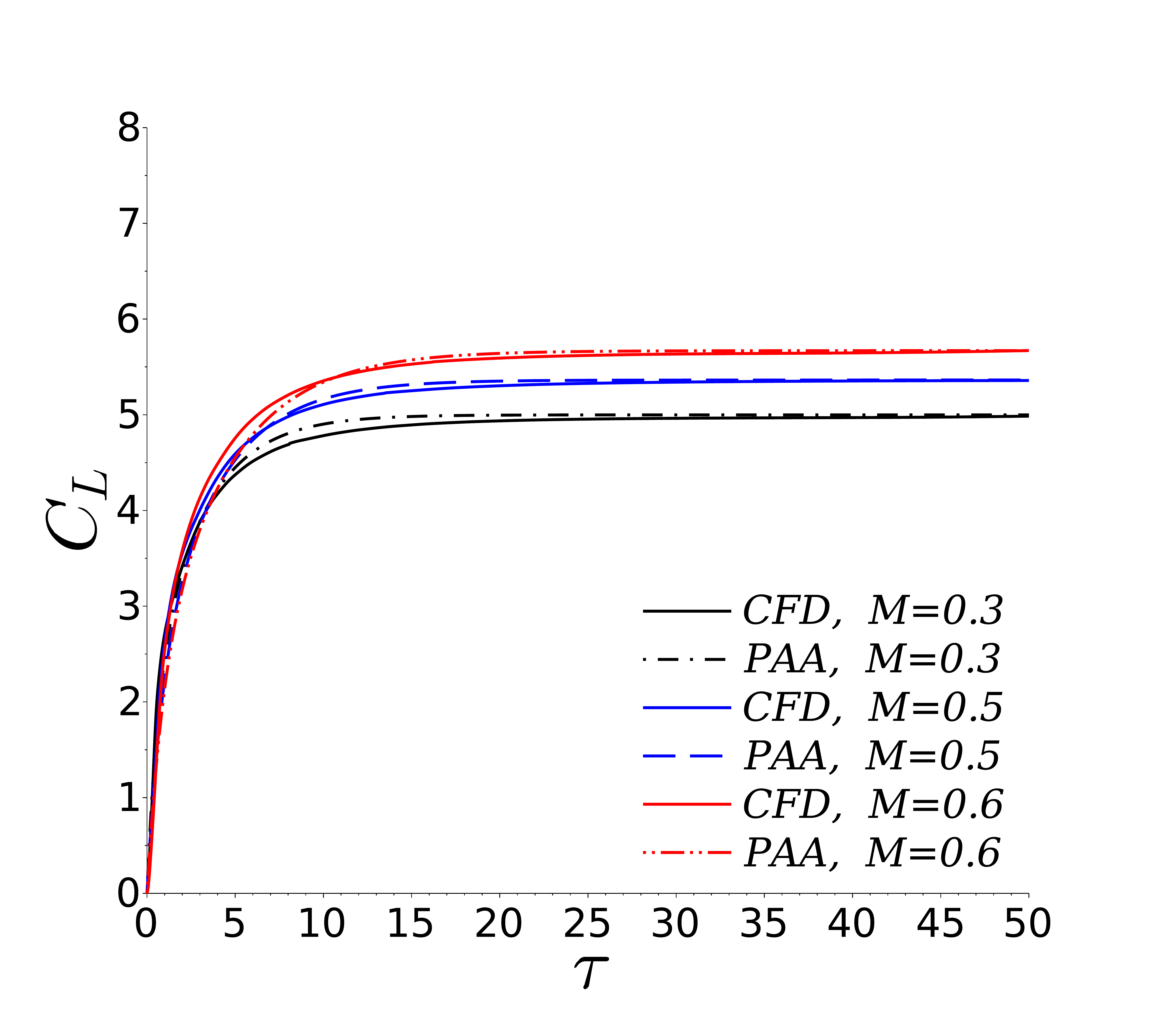}
\end{center}
\caption{Wing lift coefficient for $\eta=8$, unit AoA step (left) and unit SEG (right).}
\label{fig:aoa_ar8}
\end{figure}

\begin{figure}
\begin{center}
\includegraphics[width=0.49\textwidth]{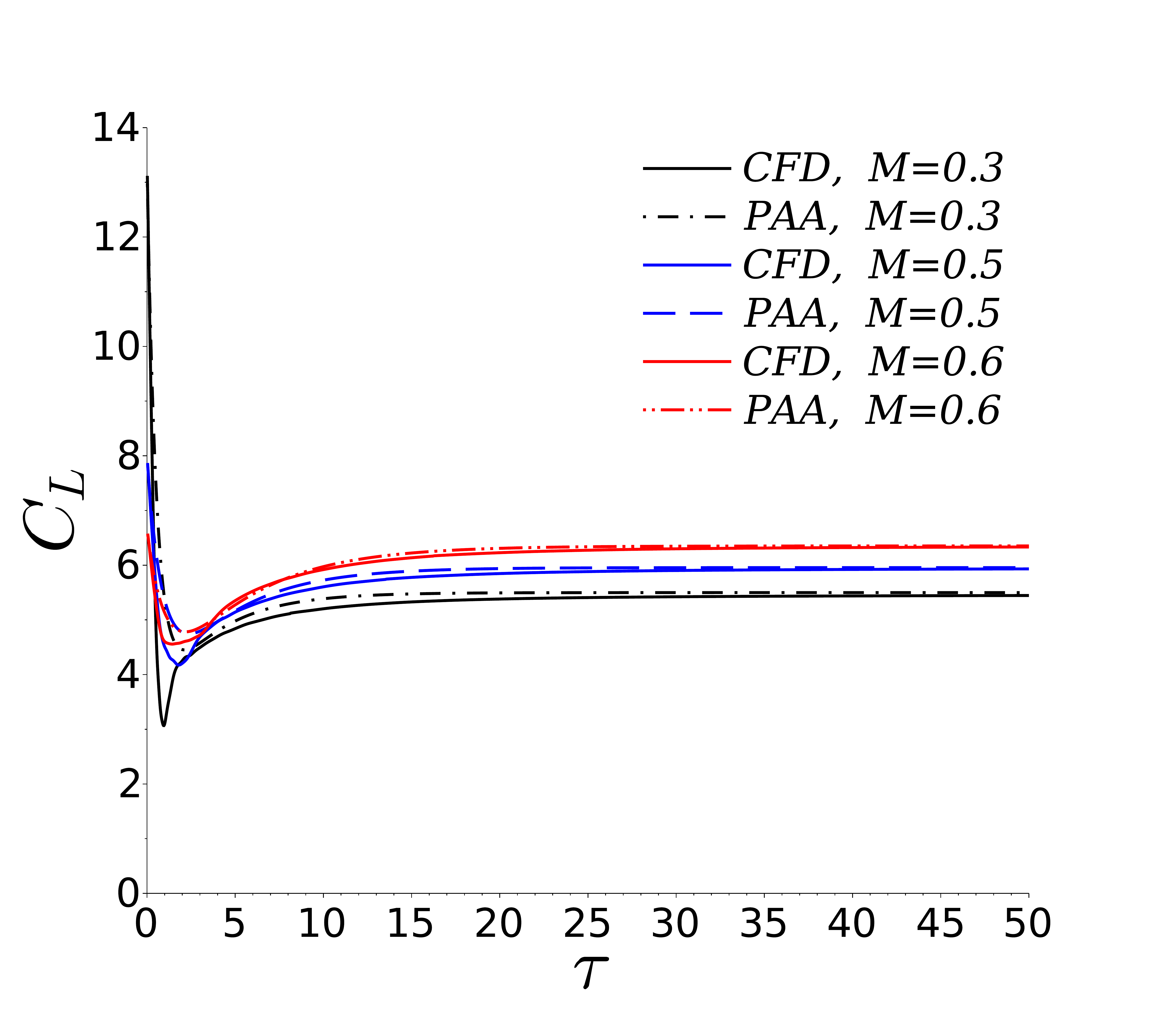}
\includegraphics[width=0.49\textwidth]{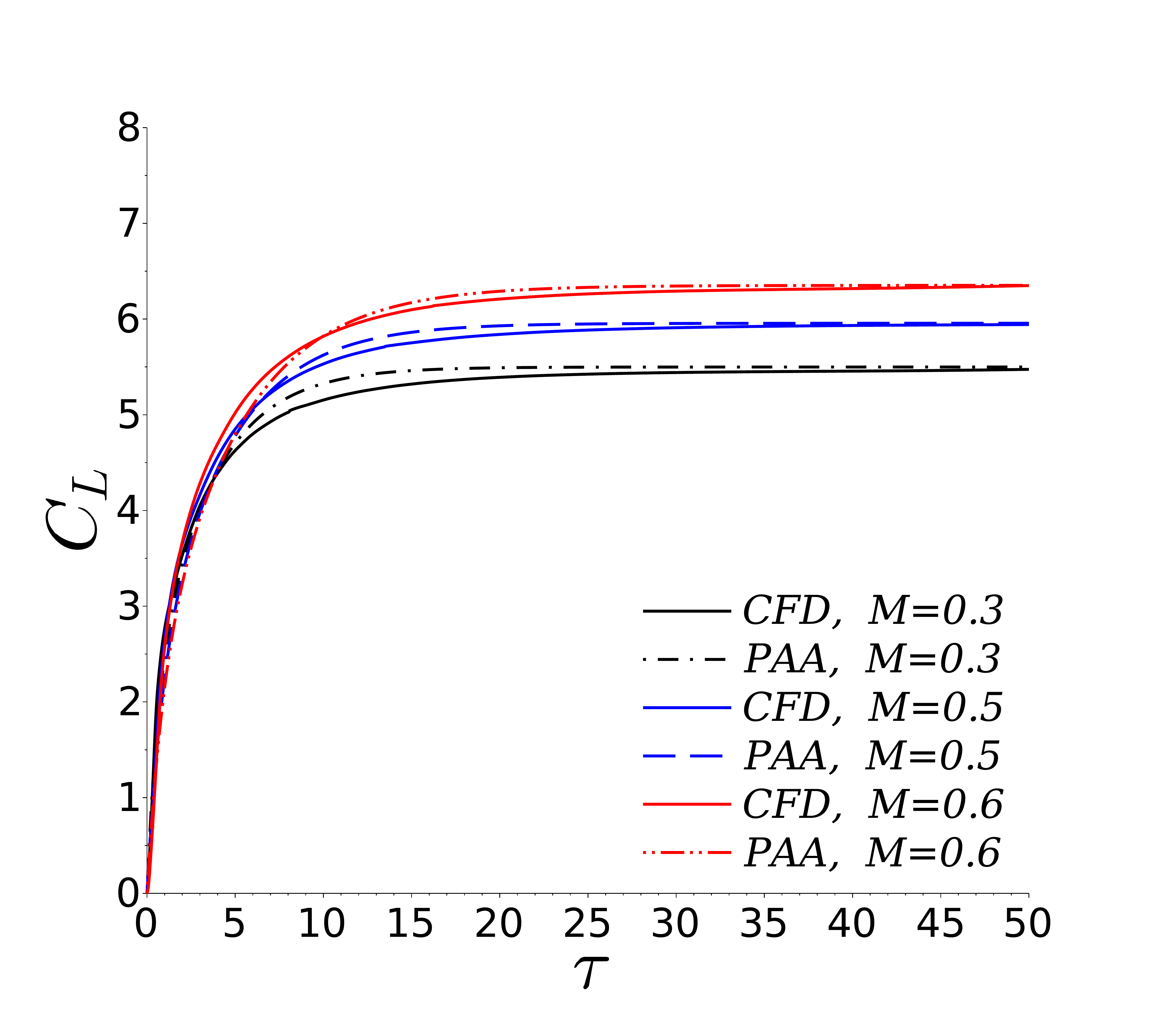}
\end{center}
\caption{Wing lift coefficient for $\eta=12$, unit AoA step (left) and unit SEG (right).}
\label{fig:aoa_ar12}
\end{figure}

\begin{figure}
\begin{center}
\includegraphics[width=0.49\textwidth]{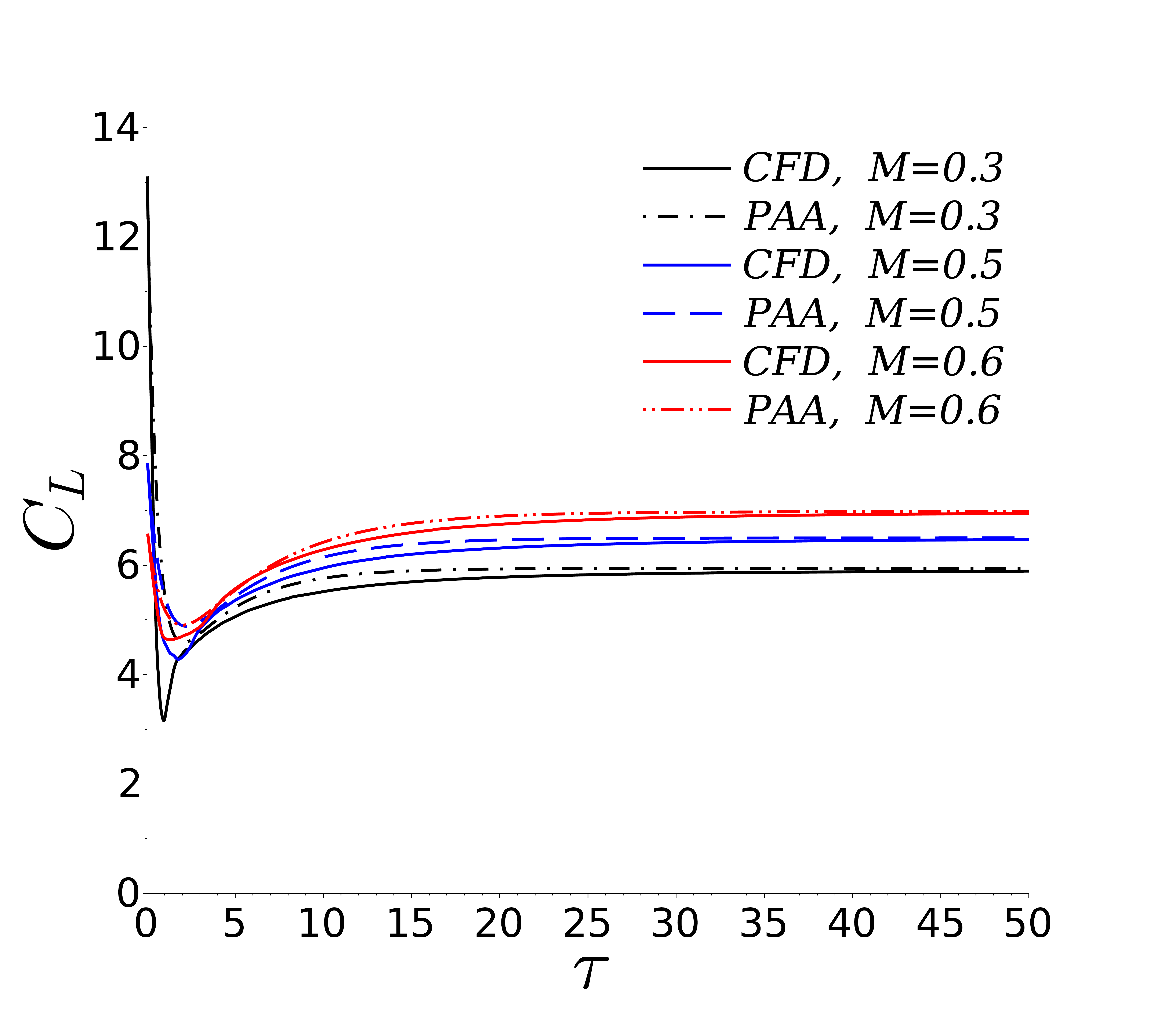}
\includegraphics[width=0.49\textwidth]{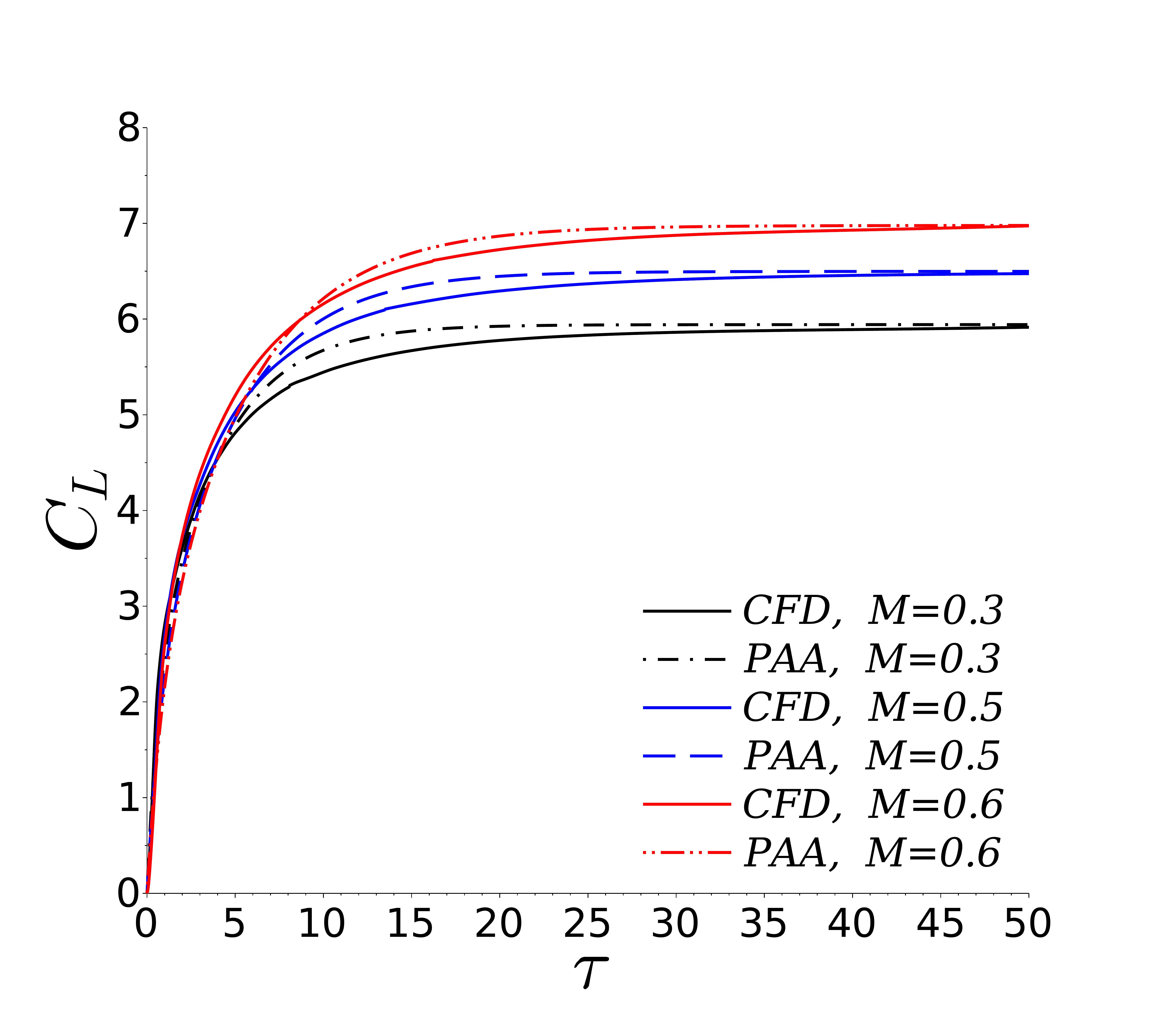}
\end{center}
\caption{Wing lift coefficient for $\eta=20$, unit AoA step (left) and unit SEG (right).}
\label{fig:aoa_ar20}
\end{figure}

The lift development of each configuration to the flow perturbations are presented in Figures \ref{fig:aoa_ar6} to \ref{fig:aoa_ar20}, for both unit AoA step and unit SEG; all Mach numbers are plotted on the same chart, but with different colours (i.e., black for $M=0.3$, blue for $M=0.5$, red for $M=0.6$) for the CFD results and different styles (i.e., dotted for $M=0.3$, dashed for $M=0.5$, dash-dot for $M=0.6$) for the parametric analytical approximations (PAA).

\subsection{Discussion}
\label{subsec:discussion}

CFD results exhibit overall good quality and consistency with each other, showing clear physical trends with the free-stream Mach number and wing aspect ratio. Numerical oscillations are virtually absent: this represents a remarkable improvement over the results obtained by the authors in the past \cite{Righi}.

The overall agreement with the parametric analytical approximations is good but in "transitional" part of the wing lift development, where the flow evolution is dominated neither by interacting pressure waves nor by the circulation build-up and straightforward analytical solutions are not available. The CFD results at lower Mach numbers and aspect ratios are closer to the analytical results than those at higher values of both parameters, since the circulatory lift build-up takes longer in the former cases and a single exponential term struggles to reproduce it (more terms and curve-fitting would then be necessary \cite{Berci-Aerospace}); the same is true at the start of a unit SEG with front normal to the reference airspeed. Note that the agreement for unit AoA step is better than that for unit SEG, since the realistic gust in the CFD simulations hits different wing sections at different times and is influenced by the wing-tip vortexes development, especially for low aspect ratios.
{
An open issue is the sensitivity of these results to parameters other than Mach number and Aspect Ratio, in particular the thickness of the wing sections, taper and sweep angle. We expect that the approach proposed here can be applied more in general, as shown in a different publication by the authors \cite{DARONCH2018617}; however, a more comprehensive study of these sensitivities will be the object of future investigations.}

\section{Conclusions}
\label{sec:conclusions}

In this study, aerodynamic indicial-admittance functions for the unsteady lift of elliptical wings in subsonic flow have been generated via CFD and approximate analytical formulations. 

Although time-accurate simulations are possible with all CFD solvers, we have investigated and assessed many factors that may prevent the users from obtaining correct results in practical cases, such as spurious effects due to grid size and resolution, time step size, numerical integration scheme, unphysical gust modelling.

We have pursued good CFD quality by eliminating spurious oscillations and checked its plausibility by introducing parametric analytical approximations based on classical and well accepted theories, which are eventually tuned with CFD results in the least "invasive" way possible within a consistent framework. 

The systematic comparison of CFD and analytical models shows an overall good agreement, in particular for a step in angle of attack of wings with small aspect ratios flying at low mach number. Thus, this study also confirms that linear aerodynamics theories based on potential flow are remarkably reliable in predicting unsteady airloads, provided the initial and asymptotic behaviours are effectively tuned with higher-fidelity data (in the case, CFD results); this multi-level strategy combines the complex generality of the numerical approach with the sound synthesis of the analytical approach, allowing a thorough understanding of the physical phenomena along with the related models applicability and cross-validation.

The added value provided by this study is associated with (i) the settings used for well-behaved CFD simulations, (ii) a method to introduce a vertical sharp-edged gust that minimises its undesired distortion as well as spurious non-physical phenomena, (iii) the development of simple parametric yet physically meaningful analytical approximations, (iv) the generation of a small database of responses which the authors are willing to share with all interested parties. This work may further be expanded in order to investigate additional aspect ratios, wing geometries, aerofoil types and Mach numbers. 

\section*{Appendix: Foundations of the Analytical Approximations}
\label{sec:appendix}

\subsection*{Non-Circulatory Compressible Flow}

At the impulsive start of the flow perturbation, the lift build-up for a unit step in the angle-of-attack $C^{\perp}_L (\tau)$ and a unit sharp-edged gust with front tangential to the leading-edge $C^{\parallel}_L (\tau)$ is linearly expressed as \cite{Lomax2,Righi}:
\begin{equation}
C^{\perp}_L = \cfrac{4}{M}\left[ 1 - \left(\cfrac{1-M}{2M}\right) \invbreve{k}^{\perp} \tau \right],  \qquad \qquad
C^{\parallel}_L = \cfrac{2 \invbreve{k}^{\parallel} \tau}{\sqrt{M}}, \qquad \tau\leq\cfrac{2M}{1+M},   
\end{equation}
respectively, within the reduced time taken by the outgoing acoustic waves to travel the aerofoil chord \cite{Lomax1}; note that penetration effects also occur while the gust travels the latter at the reference airspeed for $\tau \le 2$ \cite{Jones1}. When the gust front is normal to the reference airspeed, each wing section encounters the gust at a different time and the local delay shall be considered to calculate the lift build-up $C^{\dashv}_{L}(\tau)$ \cite{DARONCH2018617}, namely:
\begin{equation}
C^{\dashv}_L \approx
\begin{cases} 
\cfrac{4 \invbreve{k}^{\parallel} \tau^2}{\sqrt{M}},   \qquad \qquad \qquad  0\leq\tau\leq\frac{1}{4}, \\ \\
\cfrac{2 \invbreve{k}^{\parallel} \tau}{\sqrt{M}} \left(\tau-\cfrac{1}{8}\right),   \qquad   \frac{1}{4}\leq\tau\leq\frac{1}{2},
\end{cases}
\end{equation}
within the reduced time taken by $C^{\dashv}_{L}$ to grow with same rate as $C^{\parallel}_L$, following piston theory \cite{Pike}. The full wing span being impinged for $\tau\geq\frac{1}{2}$ already, the effective delay $\delta\tau=\frac{1}{8}$ is rather small and may be neglected for the initial non-circulatory load development but must be retained when deriving the tuning parameter $\invbreve{k}^{\parallel}$ from CFD simulations as well as considered for the subsequent circulatory load development.

\subsection*{Circulatory Incompressible Flow}

As for the circulatory flow contribution, with $E(\eta)$ the ratio of the wing semi-perimeter to the span \cite{Jones1,Jones3D}, the initial lift build-up for a unit step in the angle-of-attack $\breve{C}^{\perp}_L (\tau)$ and a unit sharp-edged gust with front tangential to the leading-edge $\breve{C}^{\parallel}_L (\tau)$ in incompressible flow is estimated as \cite{Jones1,Berci4}:
\begin{equation}
\breve{C}^{\perp}_L \approx \cfrac{\pi}{E} \left(1+\cfrac{\tau}{4E} \right),  \qquad \qquad
\breve{C}^{\parallel}_L \approx \cfrac{2}{E} \sqrt{2\tau},
\end{equation}
respectively, within the reduced time taken by the trailed wake (assumed as flat) to travel a few aerofoil semi-chords behind the wing; still, when the gust front is normal to the reference airspeed, the local delay shall be considered to calculate the circulatory lift build-up $\breve{C}^{\dashv}_{L}(\tau)$. Note that the circulatory flow response grows until the steady lift coefficient $\bar{C}_L$ is asymptotically reached, while the non-circulatory one decays quite rapidly.


\bibliography{berci_correct,Minhabiblioteca}

\begin{thebibliography}{10}

\bibitem{Alexandrov}
N.M. Alexandrov and M.Y. Hussaini.
\newblock {\em Multidisciplinary Design Optimization: State of the Art}.
\newblock Proceedings in Applied Mathematics Series, SIAM, 1997.

\bibitem{Basu}
B.C. Basu and G.J. Hancock.
\newblock The unsteady motion of a two-dimensional aerofoil in incompressible
  inviscid flow.
\newblock {\em Journal of Fluid Mechanics}, 87:159--178, 1978.

\bibitem{Beddoes}
T.S. Beddoes.
\newblock Practical computation of unsteady lift.
\newblock {\em Vertica}, 8(1):55--71, 1984.

\bibitem{Bellinger}
D.~Bellinger and T.~Pototzky.
\newblock A study of aerodynamic matrix numerical condition.
\newblock {\em Proceedings of the 3rd MSC Worldwide Aerospace Conference and
  Technology Showcase}, Paper 2001-21, 2001.

\bibitem{Berci4}
M.~Berci.
\newblock Lift-deficiency functions of elliptical wings in incompressible
  potential flow: Jones' theory revisited.
\newblock {\em Journal of Aircraft}, 53(2):559--602, 2016.

\bibitem{Berci-Aerospace}
M.~Berci and R.~Cavallaro.
\newblock A hybrid reduced-order model for the aeroelastic analysis of flexible
  subsonic wings - a parametric assessment.
\newblock {\em Aerospace}, 5(3):1--23, 2018.

\bibitem{Berci3}
M.~Berci, S.~Mascetti, A.~Incognito, P.H. Gaskell, and V.V. Toropov.
\newblock Dynamic response of typical section using variable-fidelity fluid
  dynamics and gust-modeling approaches - with correction methods.
\newblock {\em Journal of Aerospace Engineering}, 27(5):1--20, 2014.

\bibitem{Berci-AST-2017}
M.~Berci and M.~Righi.
\newblock An enhanced analytical method for the subsonic indicial lift of
  two-dimensional aerofoils - with numerical cross-validation.
\newblock {\em Aerospace Science and Technology}, 67:354--365, 2017.

\bibitem{Bisplinghoff}
R.L. Bisplinghoff, H.~Ashley, and R.L. Halfman.
\newblock {\em Aeroelasticity}.
\newblock Dover, 1996.

\bibitem{Cavagna2}
L.~Cavagna, G.~Quaranta, G.~Ghiringhelli, and P.~Mantegazza.
\newblock Efficient application of cfd aeroelastic methods using commercial
  software.
\newblock In {\em International forum on aeroelasticity and structural dynamics
  IFASD-2005, Munich, Germany}. Citeseer, 2005.

\bibitem{Cizmas}
P.G.A. Cizmas and J.I. Gargoloff.
\newblock Mesh generation and deformation algorithm for aeroelasticity
  simulations.
\newblock {\em Journal of Aircraft}, 45(3):1062--1066, 2008.

\bibitem{DaRonch}
A.~Da~Ronch, M.~Ghoreyshi, and K.J. Badcock.
\newblock On the generation of flight dynamics aerodynamic tables by
  computational fluid dynamics.
\newblock {\em Progress in Aerospace Sciences}, 47:597--620, 2011.

\bibitem{DARONCH2018617}
A.~Da~Ronch, A.~Ventura, M.~Righi, M.~Franciolini, M.~Berci, and D.~Kharlamov.
\newblock Extension of analytical indicial aerodynamics to generic trapezoidal
  wings in subsonic flow.
\newblock {\em Chinese Journal of Aeronautics}, 31(4):617 -- 631, 2018.

\bibitem{Diederich}
F.W. Diederich.
\newblock A plan-form parameter for correlating certain aerodynamic
  characteristics of swept wings.
\newblock {\em NACA TN 2335}, 1951.

\bibitem{economon2016su2}
T.D. Economon, F.~Palacios, S.R. Copeland, T.W. Lukaczyk, and J.J. Alonso.
\newblock {SU2}: An open-source suite for multiphysics simulation and design.
\newblock {\em AIAA Journal}, 54(3):828--846, 2016.

\bibitem{Farhat2}
C.~Farhat and V.K. Lakshminarayan.
\newblock An ale formulation of embedded boundary methods for tracking boundary
  layers in turbulent fluid-structure interaction problems.
\newblock {\em Journal of Computational Physics}, 263:53--70, 2014.

\bibitem{Farhat1}
C.~Farhat, M.~Lesoinne, and P.~LeTallec.
\newblock Load and motion transfer algorithms for fluid/structure interaction
  problems with non-matching discrete interfaces: Momentum and energy
  conservation, optimal discretization and application to aeroelasticity.
\newblock {\em Computer Methods in Applied Mechanics and Engineering},
  157(1):95--114, 1998.

\bibitem{Garrick}
L.E. Garrick.
\newblock On some reciprocal relations in the theory of nonstationary flows.
\newblock {\em NACA 629}, 1938.

\bibitem{Gennaretti}
M.~Gennaretti and F.~Mastroddi.
\newblock Study of reduced-order models for gust-response analysis of flexible
  fixed wings.
\newblock {\em Journal of Aircraft}, 41(2):304--313, 2004.

\bibitem{Ghoreyshi1}
M.~Ghoreyshi, A.~Jirasek, and R.M. Cummings.
\newblock Computational investigation into the use of response functions for
  aerodynamic-load modeling.
\newblock {\em AIAA Journal}, 50(6):1314--1327, 2012.

\bibitem{Ghoreyshi2}
M.~Ghoreyshi, A.~Jirasek, and R.M. Cummings.
\newblock Reduced order unsteady aerodynamic modeling for stability and control
  analysis using computational fluid dynamics.
\newblock {\em Progress in Aerospace Sciences}, 71:167--217, 2014.

\bibitem{Glauert2}
H.~Glauert.
\newblock The effect of compressibility on the lift of an aerofoil.
\newblock {\em Proceedings of the Royal Society of London. Series A},
  118(779):113--119, 1928.

\bibitem{Gothert}
B.H. Gothert.
\newblock Plane and three-dimensional flow at high subsonic speeds (extension
  of the prandtl rule).
\newblock {\em NACA TM-1105}, 1946.

\bibitem{Hernandes}
F.~Hernandes and P.~Soviero.
\newblock Unsteady aerodynamic coefficients obtained by a compressible vortex
  lattice method.
\newblock {\em Journal of Aircraft}, 46(4):1291--1301, 2009.

\bibitem{jameson2009}
A.~Jameson and S.~Schenectady.
\newblock An assessment of dual-time stepping, time spectral and artificial
  compressibility based numerical algorithms for unsteady flow with
  applications to flapping wings.
\newblock {\em AIAA Paper 2009-4273}, 2009.

\bibitem{Jansson}
N.~Jansson and D.~Eller.
\newblock Efficient laplace-domain aerodynamics for load analyses.
\newblock {\em IFASD-2013-000}, 2013.

\bibitem{Jones1}
R.T. Jones.
\newblock The unsteady lift of a wing of finite aspect ratio.
\newblock {\em NACA 681}, 1940.

\bibitem{Jones3D}
R.T. Jones.
\newblock Correction of the lifting-line theory for the effect of the chord.
\newblock {\em NACA TN 817}, 1941.

\bibitem{Jones2}
R.T. Jones.
\newblock Classical aerodynamic theory.
\newblock {\em NASA RP-1050}, 1979.

\bibitem{jones1945}
W.P. Jones.
\newblock Aerodynamic forces on wings in non-uniform motion.
\newblock {\em ARC-RM-2117}, 1945.

\bibitem{Jones3}
W.P. Jones.
\newblock Oscillating wings in compressible subsonic flow.
\newblock {\em ARC RM-2855}, 1957.

\bibitem{Karamcheti}
K.~Karamcheti.
\newblock {\em Principles of Ideal-Fluid Aerodynamics}.
\newblock Wiley, 1967.

\bibitem{Katz}
J.~Katz and A.~Plotkin.
\newblock {\em Low Speed Aerodynamics}.
\newblock Cambridge University Press, 2001.

\bibitem{Kier2}
T.M. Kier and G.H.N. Looye.
\newblock Unifying manoeuvre and gust loads analysis models.
\newblock {\em IFASD-2009-106}, 2009.

\bibitem{Kussner}
H.G. K{\"u}ssner.
\newblock Zusammenfassender bericht {\"u}ber den instation{\"a}ren auftrieb von
  fl{\"u}geln.
\newblock {\em Luftfahrtforschung}, 13(12):410--424, 1936.

\bibitem{Leishman1}
J.G. Leishman.
\newblock {\em Principles of Helicopter Aerodynamics}.
\newblock Cambridge Aerospace Series, Cambridge University Press, 2006.

\bibitem{Leishman4}
J.G. Leishman and K.Q. Nguyen.
\newblock State-space representation of unsteady airfoil behavior.
\newblock {\em AIAA Journal}, 28(5):836--844, 1990.

\bibitem{Lomax2}
H.~Lomax, F.D. Fuller, and L.~Sluder.
\newblock Two- and three-dimensional unsteady lift problems in high-speed
  flight.
\newblock {\em NACA 1077}, 1952.

\bibitem{Lomax1}
H.~Lomax and M.A. Heaslet.
\newblock The indicial lift and pitching moment for a sinking or pitching
  two-dimensional wing flying at subsonic or supersonic speeds.
\newblock {\em NACA TN-2403}, 1951.

\bibitem{Mazelsky1}
B.~Mazelsky.
\newblock Numerical determination of indicial lift of a two-dimensional sinking
  airfoil at subsonic mach numbers from oscillatory lift coefficients with
  calculations for mach number 0.7.
\newblock {\em NACA TN-2562}, 1951.

\bibitem{Miranda}
I.~Miranda and P.~Soviero.
\newblock Indicial response of thin wings in a compressible subsonic flow.
\newblock {\em Proceedings of COBEM 2005}, 2005.

\bibitem{Palacios}
R.~Palacios, H.~Climent, A.~Karlsson, and B.~Winzell.
\newblock Assessment of strategies for correcting linear unsteady aerodynamics
  using cfd or test results.
\newblock {\em IFASD-2001-074}, 2001.

\bibitem{Parameswaran}
V.~Parameswaran and J.~D. Baeder.
\newblock Indicial aerodynamics in compressible flow-direct computational fluid
  dynamic calculations.
\newblock {\em Journal of Aircraft}, 34(1):131--133, 1997.

\bibitem{Pike}
E.C. Pike.
\newblock Manual on aeroelasticity (volume ii, chapter vi).
\newblock {\em AGARD 578}, 1971.

\bibitem{Prandtl}
L.~Prandtl.
\newblock Applications of modern hydrodynamics to aeronautics.
\newblock {\em NACA TR-116}, 1921.

\bibitem{queijo1978approximate}
M.J. Queijo, W.R. Wells, and D.A. Keskar.
\newblock Approximate indicial lift function for tapered, swept wings in
  incompressible flow.
\newblock {\em NASA TP-1241}, 1978.

\bibitem{Raveh}
D.E. Raveh.
\newblock Reduced-order models for nonlinear unsteady aerodynamics.
\newblock {\em AIAA Journal}, 39(8):1417--1429, 2001.

\bibitem{righi2016subsonic}
M.~Righi, M.~Berci, M.~Franciolini, A.~Da~Ronch, and D.~Kharlamov.
\newblock Subsonic indicial aerodynamics for the unsteady loads of trapezoidal
  wings.
\newblock In {\em 34th AIAA Applied Aerodynamics Conference}, page 4165, 2016.

\bibitem{Righi}
M.~Righi, J.~Koch, and M.~Berci.
\newblock Subsonic indicial aerodynamics for unsteady loads calculation via
  numerical and analytical methods: a preliminary assessment.
\newblock {\em AIAA-2015-3170}, 2015.

\bibitem{Romanelli}
G.~Romanelli, E.~Serioli, and P.~Mantegazza.
\newblock A "free" approach to computational aeroelasticity.
\newblock {\em AIAA-2010-176}, 2010.

\bibitem{Silva}
W.~Silva.
\newblock Discrete-time linear and nonlinear aerodynamic impulse responses for
  efficient use of cfd analyses.
\newblock {\em PhD Thesis, College of William and Mary}, 1997.

\bibitem{Venkatesan}
C.~Venkatesan and P.~Friedmann.
\newblock New approach to finite-state modeling of unsteady aerodynamics.
\newblock {\em AIAA Journal}, 24(12):1889--1897, 1986.

\bibitem{VonKarman}
T.~Von~Karman and W.R. Sears.
\newblock Airfoil theory for non-uniform motion.
\newblock {\em Journal of the Aeronautical Sciences}, 5(10):379--390, 1938.

\bibitem{Wagner}
H.~Wagner.
\newblock {\"U}ber die entstehung des dynamischen auftriebes von
  tragfl{\"u}geln.
\newblock {\em ZAMM-Journal of Applied Mathematics and Mechanics/Zeitschrift
  f{\"u}r Angewandte Mathematik und Mechanik}, 5(1):17--35, 1925.

\bibitem{Wieseman}
C.D. Wieseman.
\newblock Methodology for matching experimental and computational aerodynamic
  data.
\newblock {\em NASA TM-100592}, 1988.

\bibitem{Wright}
J.R. Wright and J.E. Cooper.
\newblock {\em Introduction to Aircraft Aeroelasticity and Loads}.
\newblock AIAA Education Series, AIAA, 2007.

\end{thebibliography}
\bibliographystyle{plain}

\end{document}